\documentclass[twocolumn,letterpaper,aps,prd,superscriptaddress,showpacs,floatfix,longbibliography]{revtex4-2}

\usepackage{graphicx}	
\usepackage{amsmath}

\usepackage{xspace}	
\emergencystretch=\maxdimen
\newcommand{\pt}{\mbox{$p_T$}\xspace}

\newcommand{\dAu}{\mbox{$d$$+$Au}\xspace}
\newcommand{\pA}{\mbox{$p$$+$$A$}\xspace}
\newcommand{\pp}{\mbox{$p$$+$$p$}\xspace}
\newcommand{\pAu}{\mbox{$p$$+$Au}\xspace}
\newcommand{\pAl}{\mbox{$p$$+$Al}\xspace}
\newcommand{\polpp}{\mbox{$p^{\uparrow}$$+$$p$}\xspace}
\newcommand{\polpA}{\mbox{$p^{\uparrow}$$+$$A$}\xspace}
\newcommand{\polpAu}{\mbox{$p^{\uparrow}$$+$Au}\xspace}
\newcommand{\polpAl}{\mbox{$p^{\uparrow}$$+$Al}\xspace}
\newcommand{\absetarange}{\mbox{$1.2<|\eta|<2.4$}\xspace}

\begin{document}

\title{Transverse single-spin asymmetry of charged hadrons at forward 
and backward rapidity in polarized $p$$+$$p$, $p$$+$Al, and $p$$+$Au collisions 
at $\sqrt{s_{_{NN}}}=200$ GeV}


\newcommand{\abilene}{Abilene Christian University, Abilene, Texas 79699, USA}
\newcommand{\augie}{Department of Physics, Augustana University, Sioux Falls, South Dakota 57197, USA}
\newcommand{\banaras}{Department of Physics, Banaras Hindu University, Varanasi 221005, India}
\newcommand{\barc}{Bhabha Atomic Research Centre, Bombay 400 085, India}
\newcommand{\baruch}{Baruch College, City University of New York, New York, New York, 10010 USA}
\newcommand{\bnlcoll}{Collider-Accelerator Department, Brookhaven National Laboratory, Upton, New York 11973-5000, USA}
\newcommand{\bnlphys}{Physics Department, Brookhaven National Laboratory, Upton, New York 11973-5000, USA}
\newcommand{\caucr}{University of California-Riverside, Riverside, California 92521, USA}
\newcommand{\charlesczech}{Charles University, Faculty of Mathematics and Physics, 180 00 Troja, Prague, Czech Republic}
\newcommand{\cns}{Center for Nuclear Study, Graduate School of Science, University of Tokyo, 7-3-1 Hongo, Bunkyo, Tokyo 113-0033, Japan}
\newcommand{\colorado}{University of Colorado, Boulder, Colorado 80309, USA}
\newcommand{\columbia}{Columbia University, New York, New York 10027 and Nevis Laboratories, Irvington, New York 10533, USA}
\newcommand{\czechtech}{Czech Technical University, Zikova 4, 166 36 Prague 6, Czech Republic}
\newcommand{\debrecen}{Debrecen University, H-4010 Debrecen, Egyetem t{\'e}r 1, Hungary}
\newcommand{\elte}{ELTE, E{\"o}tv{\"o}s Lor{\'a}nd University, H-1117 Budapest, P{\'a}zm{\'a}ny P.~s.~1/A, Hungary}
\newcommand{\ewha}{Ewha Womans University, Seoul 120-750, Korea}
\newcommand{\famu}{Florida A\&M University, Tallahassee, FL 32307, USA}
\newcommand{\fsu}{Florida State University, Tallahassee, Florida 32306, USA}
\newcommand{\gsu}{Georgia State University, Atlanta, Georgia 30303, USA}
\newcommand{\hiroshima}{Physics Program and International Institute for Sustainability with Knotted Chiral Meta Matter (SKCM2), Hiroshima University, Higashi-Hiroshima, Hiroshima 739-8526, Japan}
\newcommand{\howard}{Department of Physics and Astronomy, Howard University, Washington, DC 20059, USA}
\newcommand{\ihepprot}{IHEP Protvino, State Research Center of Russian Federation, Institute for High Energy Physics, Protvino, 142281, Russia}
\newcommand{\illuiuc}{University of Illinois at Urbana-Champaign, Urbana, Illinois 61801, USA}
\newcommand{\inrras}{Institute for Nuclear Research of the Russian Academy of Sciences, prospekt 60-letiya Oktyabrya 7a, Moscow 117312, Russia}
\newcommand{\instpasczech}{Institute of Physics, Academy of Sciences of the Czech Republic, Na Slovance 2, 182 21 Prague 8, Czech Republic}
\newcommand{\isu}{Iowa State University, Ames, Iowa 50011, USA}
\newcommand{\jaea}{Advanced Science Research Center, Japan Atomic Energy Agency, 2-4 Shirakata Shirane, Tokai-mura, Naka-gun, Ibaraki-ken 319-1195, Japan}
\newcommand{\jeonbuk}{Jeonbuk National University, Jeonju, 54896, Korea}
\newcommand{\kek}{KEK, High Energy Accelerator Research Organization, Tsukuba, Ibaraki 305-0801, Japan}
\newcommand{\korea}{Korea University, Seoul 02841, Korea}
\newcommand{\kurchatov}{National Research Center ``Kurchatov Institute", Moscow, 123098 Russia}
\newcommand{\kyoto}{Kyoto University, Kyoto 606-8502, Japan}
\newcommand{\lawllnl}{Lawrence Livermore National Laboratory, Livermore, California 94550, USA}
\newcommand{\losalamos}{Los Alamos National Laboratory, Los Alamos, New Mexico 87545, USA}
\newcommand{\lund}{Department of Physics, Lund University, Box 118, SE-221 00 Lund, Sweden}
\newcommand{\lyon}{IPNL, CNRS/IN2P3, Univ Lyon, Universit{\'e} Lyon 1, F-69622, Villeurbanne, France}
\newcommand{\maryland}{University of Maryland, College Park, Maryland 20742, USA}
\newcommand{\mass}{Department of Physics, University of Massachusetts, Amherst, Massachusetts 01003-9337, USA}
\newcommand{\mate}{MATE, Laboratory of Femtoscopy, K\'aroly R\'obert Campus, H-3200 Gy\"ongy\"os, M\'atrai\'ut 36, Hungary}
\newcommand{\michigan}{Department of Physics, University of Michigan, Ann Arbor, Michigan 48109-1040, USA}
\newcommand{\miss}{Mississippi State University, Mississippi State, Mississippi 39762, USA}
\newcommand{\muhlenberg}{Muhlenberg College, Allentown, Pennsylvania 18104-5586, USA}
\newcommand{\nara}{Nara Women's University, Kita-uoya Nishi-machi Nara 630-8506, Japan}
\newcommand{\natmephi}{National Research Nuclear University, MEPhI, Moscow Engineering Physics Institute, Moscow, 115409, Russia}
\newcommand{\newmex}{University of New Mexico, Albuquerque, New Mexico 87131, USA}
\newcommand{\nmsu}{New Mexico State University, Las Cruces, New Mexico 88003, USA}
\newcommand{\northcg}{Physics and Astronomy Department, University of North Carolina at Greensboro, Greensboro, North Carolina 27412, USA}
\newcommand{\ohio}{Department of Physics and Astronomy, Ohio University, Athens, Ohio 45701, USA}
\newcommand{\ornl}{Oak Ridge National Laboratory, Oak Ridge, Tennessee 37831, USA}
\newcommand{\orsay}{IPN-Orsay, Univ.~Paris-Sud, CNRS/IN2P3, Universit\'e Paris-Saclay, BP1, F-91406, Orsay, France}
\newcommand{\peking}{Peking University, Beijing 100871, People's Republic of China}
\newcommand{\pnpi}{PNPI, Petersburg Nuclear Physics Institute, Gatchina, Leningrad region, 188300, Russia}
\newcommand{\pusan}{Pusan National University, Pusan 46241, Korea}
\newcommand{\riken}{RIKEN Nishina Center for Accelerator-Based Science, Wako, Saitama 351-0198, Japan}
\newcommand{\rikjrbrc}{RIKEN BNL Research Center, Brookhaven National Laboratory, Upton, New York 11973-5000, USA}
\newcommand{\rikkyo}{Physics Department, Rikkyo University, 3-34-1 Nishi-Ikebukuro, Toshima, Tokyo 171-8501, Japan}
\newcommand{\saispbstu}{Saint Petersburg State Polytechnic University, St.~Petersburg, 195251 Russia}
\newcommand{\seoulnat}{Department of Physics and Astronomy, Seoul National University, Seoul 151-742, Korea}
\newcommand{\stonybrkc}{Chemistry Department, Stony Brook University, SUNY, Stony Brook, New York 11794-3400, USA}
\newcommand{\stonycrkp}{Department of Physics and Astronomy, Stony Brook University, SUNY, Stony Brook, New York 11794-3800, USA}
\newcommand{\tenn}{University of Tennessee, Knoxville, Tennessee 37996, USA}
\newcommand{\texsu}{Texas Southern University, Houston, TX 77004, USA}
\newcommand{\titech}{Department of Physics, Tokyo Institute of Technology, Oh-okayama, Meguro, Tokyo 152-8551, Japan}
\newcommand{\tsukuba}{Tomonaga Center for the History of the Universe, University of Tsukuba, Tsukuba, Ibaraki 305, Japan}
\newcommand{\vandy}{Vanderbilt University, Nashville, Tennessee 37235, USA}
\newcommand{\weizmann}{Weizmann Institute, Rehovot 76100, Israel}
\newcommand{\wigner}{Institute for Particle and Nuclear Physics, Wigner Research Centre for Physics, Hungarian Academy of Sciences (Wigner RCP, RMKI) H-1525 Budapest 114, POBox 49, Budapest, Hungary}
\newcommand{\yonsei}{Yonsei University, IPAP, Seoul 120-749, Korea}
\newcommand{\zagreb}{Department of Physics, Faculty of Science, University of Zagreb, Bijeni\v{c}ka c.~32 HR-10002 Zagreb, Croatia}
\newcommand{\zambia}{Department of Physics, School of Natural Sciences, University of Zambia, Great East Road Campus, Box 32379, Lusaka, Zambia}
\affiliation{\abilene}
\affiliation{\augie}
\affiliation{\banaras}
\affiliation{\barc}
\affiliation{\baruch}
\affiliation{\bnlcoll}
\affiliation{\bnlphys}
\affiliation{\caucr}
\affiliation{\charlesczech}
\affiliation{\cns}
\affiliation{\colorado}
\affiliation{\columbia}
\affiliation{\czechtech}
\affiliation{\debrecen}
\affiliation{\elte}
\affiliation{\ewha}
\affiliation{\famu}
\affiliation{\fsu}
\affiliation{\gsu}
\affiliation{\hiroshima}
\affiliation{\howard}
\affiliation{\ihepprot}
\affiliation{\illuiuc}
\affiliation{\inrras}
\affiliation{\instpasczech}
\affiliation{\isu}
\affiliation{\jaea}
\affiliation{\jeonbuk}
\affiliation{\kek}
\affiliation{\korea}
\affiliation{\kurchatov}
\affiliation{\kyoto}
\affiliation{\lawllnl}
\affiliation{\losalamos}
\affiliation{\lund}
\affiliation{\lyon}
\affiliation{\maryland}
\affiliation{\mass}
\affiliation{\mate}
\affiliation{\michigan}
\affiliation{\miss}
\affiliation{\muhlenberg}
\affiliation{\nara}
\affiliation{\natmephi}
\affiliation{\newmex}
\affiliation{\nmsu}
\affiliation{\northcg}
\affiliation{\ohio}
\affiliation{\ornl}
\affiliation{\orsay}
\affiliation{\peking}
\affiliation{\pnpi}
\affiliation{\pusan}
\affiliation{\riken}
\affiliation{\rikjrbrc}
\affiliation{\rikkyo}
\affiliation{\saispbstu}
\affiliation{\seoulnat}
\affiliation{\stonybrkc}
\affiliation{\stonycrkp}
\affiliation{\tenn}
\affiliation{\texsu}
\affiliation{\titech}
\affiliation{\tsukuba}
\affiliation{\vandy}
\affiliation{\weizmann}
\affiliation{\wigner}
\affiliation{\yonsei}
\affiliation{\zagreb}
\affiliation{\zambia}
\author{N.J.~Abdulameer} \affiliation{\debrecen}
\author{U.~Acharya} \affiliation{\gsu} 
\author{C.~Aidala} \affiliation{\michigan} 
\author{Y.~Akiba} \email[PHENIX Spokesperson: ]{akiba@rcf.rhic.bnl.gov} \affiliation{\riken} \affiliation{\rikjrbrc} 
\author{M.~Alfred} \affiliation{\howard} 
\author{V.~Andrieux} \affiliation{\michigan} 
\author{N.~Apadula} \affiliation{\isu} 
\author{H.~Asano} \affiliation{\kyoto} \affiliation{\riken} 
\author{B.~Azmoun} \affiliation{\bnlphys} 
\author{V.~Babintsev} \affiliation{\ihepprot} 
% ...
\collaboration{PHENIX Collaboration}  \noaffiliation

\date{\today}


\begin{abstract}


Reported here are transverse single-spin asymmetries ($A_{N}$) in the 
production of charged hadrons as a function of transverse momentum 
($p_T$) and Feynman-$x$ ($x_F$) in polarized $p^{\uparrow}$$+$$p$, 
$p^{\uparrow}$$+$Al, and $p^{\uparrow}$$+$Au collisions at 
$\sqrt{s_{_{NN}}}=200$ GeV. The measurements have been performed at 
forward and backward rapidity ($1.4<|\eta|<2.4$) over the range of 
$1.5<p_{T}<7.0~{\rm GeV}/c$ and $0.04<|x_{F}|<0.2$. A nonzero asymmetry 
is observed for positively charged hadrons at forward rapidity ($x_F>0$) 
in $p^{\uparrow}$$+$$p$ collisions, whereas the $p^{\uparrow}$$+$Al and 
$p^{\uparrow}$$+$Au results show smaller asymmetries. This finding 
provides new opportunities to investigate the origin of transverse 
single-spin asymmetries and a tool to study nuclear effects in $p$$+$$A$ 
collisions.

\end{abstract}


\maketitle

\section{Introduction}

An understanding of transverse single-spin asymmetries (TSSAs) in 
transversely polarized proton-proton collisions (\polpp) is crucial for 
disentangling the spin structure of the proton and parton dynamics 
within the proton. The analyzing power $A_{N}$ is defined as the 
left-right asymmetry of the produced hadrons with respect to the spin 
direction of the polarized proton where the polarization direction is 
perpendicular to the beam direction. Since the 1970s, significant TSSAs 
in hadron ($h$) production ($p^{\uparrow}$$+$$p{\rightarrow}h$$+$$X$) have been 
measured at large Feynman-$x$ ($x_{F} = 2 p_{L}/\sqrt{s}$) for a wide 
range of collision energies up to 500 
GeV~\cite{Klem:1976ui,Antille:1980th,FNAL-E704:1991ovg,E581:1991eys,Allgower:2002qi,STAR:2003lxu,Lee:2007zzh,STAR:2008ixi,BRAHMS:2008doi,STAR:2012ljf,PHENIX:2013wle,PHENIX:2014qwb,STAR:2020nnl}. 
Two approaches have been proposed to account for these large 
asymmetries. First, the transverse-momentum dependent (TMD) approach is 
based on TMD parton distribution and fragmentation functions. It 
requires two scales, a hard-scattering energy scale $Q$ and a soft scale 
$k_T$ describing the transverse momentum of partons in the proton or of 
hadrons relative to the parent parton in hadronization process, with 
$k_T \ll Q$. In this framework, the possible origins of the asymmetry 
are the Sivers~\cite{Sivers:1989cc,Sivers:1990fh} and 
Collins~\cite{Collins:1992kk} mechanisms. The Sivers mechanism describes 
the initial-state correlation between the spin of the transversely 
polarized proton and the parton transverse momentum, while in the final 
state the Collins mechanism introduces a correlation between the 
transverse spin of the fragmenting quark and transverse momentum of the 
final state hadron.  The Collins mechanism convolves with the quark 
transversity distribution, which describes the quark transverse 
polarization inside a transversely polarized proton.  Secondly, in the 
other framework, the collinear twist-3 factorization is applicable for 
observables with one large scale $Q$, which is usually represented by 
the particle transverse momentum $p_T$ in reactions such as 
$p^{\uparrow}$$+$$p{\rightarrow}h$$+$$X$. In this framework, the asymmetry arises 
from a twist-3 multi-parton correlation function and a twist-3 
fragmentation function. Twist-3 multi-parton correlation functions 
represent a spin dependence of the transverse motion of the parton inside 
a polarized 
proton~\cite{Qiu:1998ia,Kouvaris:2006zy,Koike:2007rq,Koike:2009ge,Kanazawa:2010au,Kang:2011hk,Kanazawa:2011bg,Kang:2012xf,Beppu:2013uda}. 
Twist-3 fragmentation functions denote parton fragmentation effects 
during the formation of the final state hadrons. Recent calculations of 
the twist-3 contribution from parton fragmentation are shown to be 
important to describe the Relativistic Heavy Ion Collider (RHIC) 
results~\cite{Metz:2012ct,Kanazawa:2014dca,Gamberg:2017gle}. Also, a 
recent phenomenological study using both TMD and collinear twist-3 
approaches demonstrates that TSSAs in Semi-Inclusive Deep Inelastic 
Scattering (SIDIS), Drell-Yan, $e^{+}e^{-}$ annihilation, and 
proton-proton collisions have a common origin~\cite{Cammarota:2020qcw}.

Recently, an exploration of the interplay of spin physics and small-$x$ 
physics through the measurement of TSSAs in transversely polarized 
proton-nucleus collisions (\polpA) has attracted attention, where $x$ is 
the momentum fraction of a proton carried by the parton. In \pA 
collisions, the properties of small-$x$ gluons inside nuclei can be 
probed by measuring hadron production in the proton-going direction. The 
gluon density in the small-$x$ region of the target nuclei is expected 
to increase significantly and it may be described by the 
color-glass-condensate (CGC) formalism~\cite{Gelis:2010nm}. The 
framework introduces a characteristic saturation scale $Q_{s}$ which 
describes the color-charge density fluctuations, where $Q_{sA}\propto 
A^{1/3}$ for the target nucleus. In \polpA collisions, measuring TSSAs 
can be used as a probe for the saturation scale in the 
nucleus~\cite{Kang:2011ni}. In addition, TSSAs in the forward region 
from \polpA collisions help to disentangle differing mechanisms and 
clarify the origin of the TSSA. Previous calculations of TSSAs 
incorporating gluon saturation at small-$x$ suggest that TSSAs in \polpA 
may or may not be $A$-dependent, depending on the mechanisms 
involved~\cite{Kang:2011ni, Kovchegov:2012ga, Schafer:2014zea, 
Zhou:2015ima, Hatta:2016wjz, Hatta:2016khv}. On the experimental side, 
the $A_{N}$ of positively charged hadrons at $0.1<x_{F}<0.2$ in 
PHENIX~\cite{PHENIX:2019ouo} and $\pi^{0}$ results at $0.2<x_{F}<0.7$ in 
STAR in \polpp and \polpA collisions have been reported 
recently~\cite{STAR:2020grs}. It should be noted that the kinematic 
region of the measurements presented here is outside of the expected CGC 
range, but considerations on the $A$-dependence of the various 
contributions might still be relevant.

This paper reports on measurements in 2015 of the transverse single-spin 
asymmetry for the production of charged hadrons ($h^{\pm}$) over the 
range of $1.5<p_{T}<7.0$ GeV/$c$ and $0.04<|x_{F}|<0.2$ at forward and 
backward rapidity ($1.4<|\eta|<2.4$) from transversely polarized 
proton-proton (\polpp), proton-Aluminum (\polpAl), and proton-Gold 
(\polpAu) collisions at $\sqrt{s_{_{NN}}}=200$ GeV.
Section~\ref{sec:expsetup} describes the experimental setup in 
PHENIX and the polarized beams at RHIC. The details of the analysis 
procedure are presented in Sec.~\ref{sec:analysis}, and the result are 
shown in Sec.~\ref{sec:results} with discussions in 
Sec.~\ref{sec:discussion} and finally, the summary in Sec.~\ref{sec:summary}.

\section{Experimental Setup}
\label{sec:expsetup}

\subsection{The PHENIX experiment}

\begin{figure}[thb]
\centering
\includegraphics[width=1.0\linewidth]{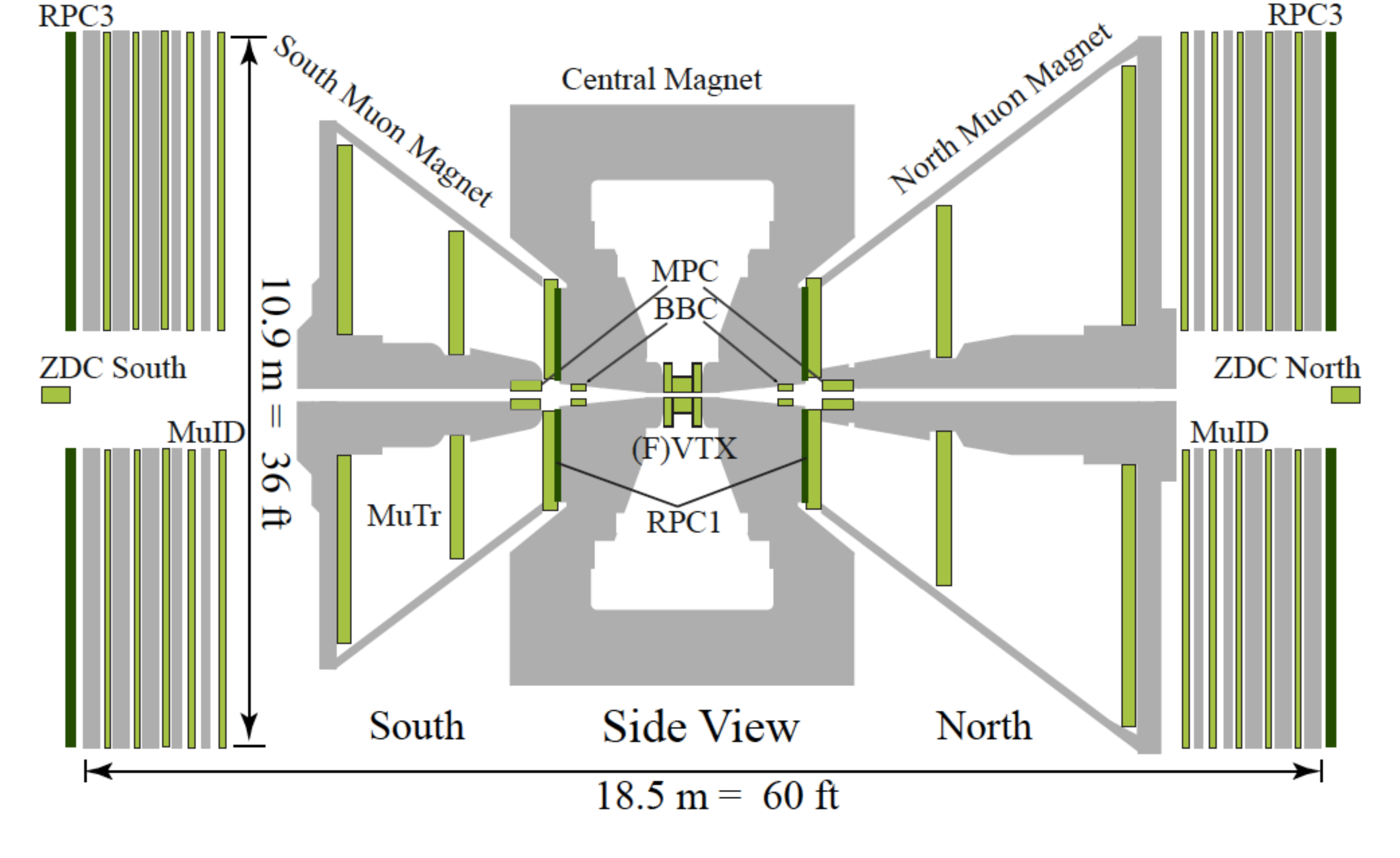}
\caption{\label{fig:PHENIX}
Side view of the PHENIX detector in 2015.}
\end{figure}

The PHENIX detector~\cite{PHENIX:2003nhg} is equipped with central arms 
at midrapidity, and muon arms at backward and forward rapidity. The 
muon arms cover the pseudorapidity range $1.2<\eta<2.4$ (north arm) and 
$-2.2<\eta<-1.2$ (south arm) and the full azimuthal angle ($\Delta \phi 
= 2\pi$)~\cite{PHENIX:2003yhi}. The side view of the PHENIX detector 
including muon arms during the 2015 run is shown in 
Fig.~\ref{fig:PHENIX}. Muons and hadrons from the interaction 
region pass through a hadron absorber of 7.5 nuclear interaction lengths 
($\lambda_{I}$). Surviving muons and hadrons reach a Muon Tracker (MuTr) 
composed of three stations of cathode-strip readout tracking chambers 
mounted inside conical-shaped muon magnets. The momentum of a charged 
particle is measured with the MuTr. The Muon Identifier (MuID) behind 
the MuTr is composed of five layers (labeled from Gap0 to Gap4) of 
proportional tube planes combined with an absorber plane 
($\approx1\lambda_{I}$), respectively. The MuID provides identification of 
hadrons and muons by measuring penetration depth. Muons of momentum 
larger than 3 GeV/$c$ penetrate all layers while charged hadrons are 
mostly stopped at the intermediate layers (Gap2 and 
Gap3)~\cite{PHENIX:2012itj}.

The Beam-Beam Counters (BBC) are located at $z=\pm144~{\rm cm}$ from the 
interaction point and cover the pseudorapidity range $3.1<|\eta|<3.9$ 
and full azimuthal angle~\cite{PHENIX:2003tlh}. Each BBC comprises 64 
quartz \v{C}erenkov counters. The BBC detects charged particles and 
provides $z$-vertex position with a resolution of $\approx$2~cm in \pp 
collisions. The BBC is also used to categorize events with energy 
deposition in terms of centrality for collisions with ion beam. In the 
case of \polpAu and \polpAl collisions, energy deposition in the 
$A$-going direction is used for determining centrality. In addition, the 
BBC serves as a luminosity monitor.

The minimum-bias trigger provided by the BBC requires at least one 
hit in both directions. The BBC trigger efficiency is 55\% in \polpp, 
72\% in \polpAl, and 84\% in \polpAu collisions. The MuID provides a 
trigger to select events containing hadron or muon tracks by requiring 
at least one MuID track reaching Gap2, Gap3 or Gap4. For charged 
hadrons, events with track candidates having a hit at the Gap2 or Gap3 
but no hit at the Gap4 are selected. The new readout electronics for the 
MuTr allow to trigger events with high momentum ($p_T>3~{\rm GeV}/c$) 
tracks by requiring maximum bending at the middle plane less than 3 MuTr 
cathode strips~\cite{Adachi:2013qha}.

\subsection{RHIC polarized beams}

RHIC is the world's first and only polarized proton collider, located at 
Brookhaven National Laboratory, comprising two counter-circulating 
storage rings. In each ring, as many as 120 bunches of heavy ions or 
polarized protons can be accelerated up to 100 GeV for heavy ions and 
255 GeV for polarized protons. During the 2015 run period, data from the 
collisions of vertically polarized protons and ions (\polpp, \polpAl, 
and \polpAu) at $\sqrt{s_{_{NN}}}=200$ GeV were recorded. 
Each beam bunch of 106~ns interval has a separate polarization (up or down). 
The predetermined polarization patterns for every eight bunches were changed 
at each fill to minimize systematic effects due to time dependence of the 
detector and accelerator performance.
In the \polpA run, the counterclockwise direction was selected for ion beams 
where the beam points towards the south muon arm.

The average polarization of the proton beam was 58\% (clockwise beam) 
and 57\% (counterclockwise beam) for \polpp collisions, 58\% for \polpAl 
collisions, and 61\% for \polpAu collisions with 3\% relative 
uncertainty from the polarization normalization~\cite{polarimetry}.

\section{Data Analysis}
\label{sec:analysis}
\subsection{Data Set}

The integrated luminosity of the data in this analysis is 
37~${\rm pb^{-1}}$ from \polpp, 593~${\rm nb^{-1}}$ from \polpAl, 
and 112~${\rm nb^{-1}}$ from \polpAu collisions. The data was 
recorded using the hadron trigger in combination with the BBC trigger. 
The hadron trigger selected high-momentum tracks when the last MuID 
gap for a track pass through is MuID Gap2 or Gap3, with the bending 
at the middle plane of MuTr being less than 3 cathode 
strips~\cite{Adachi:2013qha}.

\subsection{Charged Hadron Selection}

Among the reconstructed tracks stopped at MuID Gap2 or Gap3 are low 
momentum muons from light hadron decays. From full {\sc geant4} 
simulation studies, such background is significantly suppressed ($<5\%$) 
after applying a cut $p_z>3.5 {\rm GeV}/c$~\cite{PHENIX:2019gix}. The 
additional track quality cuts are listed in 
Table~\ref{tab:qualitycuts} with numerical values shown for Gap3 tracks 
at $1.25<p_{T}<1.5~{\rm GeV}/c$. The distance and angular difference 
between the extrapolated MuTr track and the MuID track at the first MuID 
plane's $z$ position are called DG0 and DDG0. The distance between the 
interaction point and a projected position of a MuID track at $z=0$ is 
$r_{\rm ref}$. The number of hits in a MuTr track and $\chi^{2}$ per 
degree of freedom ($\chi^{2}_{\rm MuTr}/ndf$) are also included in the 
track quality cut. The quality cuts are $p_T$ dependent except for the 
number of hits in a MuTr track and $\chi^{2}_{\rm MuTr}/ndf$. 
The polar scattering angle of a track inside the 
 absorber scaled by the momentum is $p\cdot(\theta_{\rm MuTr} - 
\theta_{\rm vtx})$. $\theta_{\rm vtx}$ is the polar angle at the collision 
vertex with the momentum vector at the vertex, obtained by a track fit from the 
MuTr and MuID to the primary vertex. $\theta_{\rm MuTr}$ is the polar angle 
at the MuTr Station 1 of the reconstructed track using the MuTr so that the 
$p\cdot(\theta_{\rm MuTr} - \theta_{\rm vtx})$ corresponds to the polar 
scattering angle of a track inside the absorber scaled by the momentum. 
Cuts applied 
to $p\cdot(\theta_{\rm MuTr} - \theta_{\rm vtx})$ and $\chi^{2}$ at 
$z_{\rm vtx}$ are effective for rejecting muons decayed inside the 
absorber.

\begin{table}[tbh]
\caption{\label{tab:qualitycuts}
Track selection cuts for tracks stopped at MuID Gap3 for a \pt bin 
$1.25<p_{T}<1.5~{\rm GeV}/c$.}
\begin{ruledtabular}
\begin{tabular}{c}
DG0 $<29~{\rm cm}$ (south), 18.5~{\rm cm} (north)\\
DDG0 $<12~{\rm deg.}$\\
$r_{\rm ref}<140~{\rm cm}$ (south), 159~{\rm cm} (north)\\
Number of hits in MuTr $>10$, $\chi^{2}_{\rm MuTr}/ndf < 20$ \\
$p\cdot(\theta_{\rm MuTr} - \theta_{\rm vtx})<0.35~{\rm rad}\cdot{\rm GeV}/c$\\
$\chi^{2}$ of track projection to $z_{\rm vtx}<8.5$
\end{tabular}
\end{ruledtabular}
\end{table}

\begin{figure}[thb]
\centering
\includegraphics[width=1.0\linewidth]{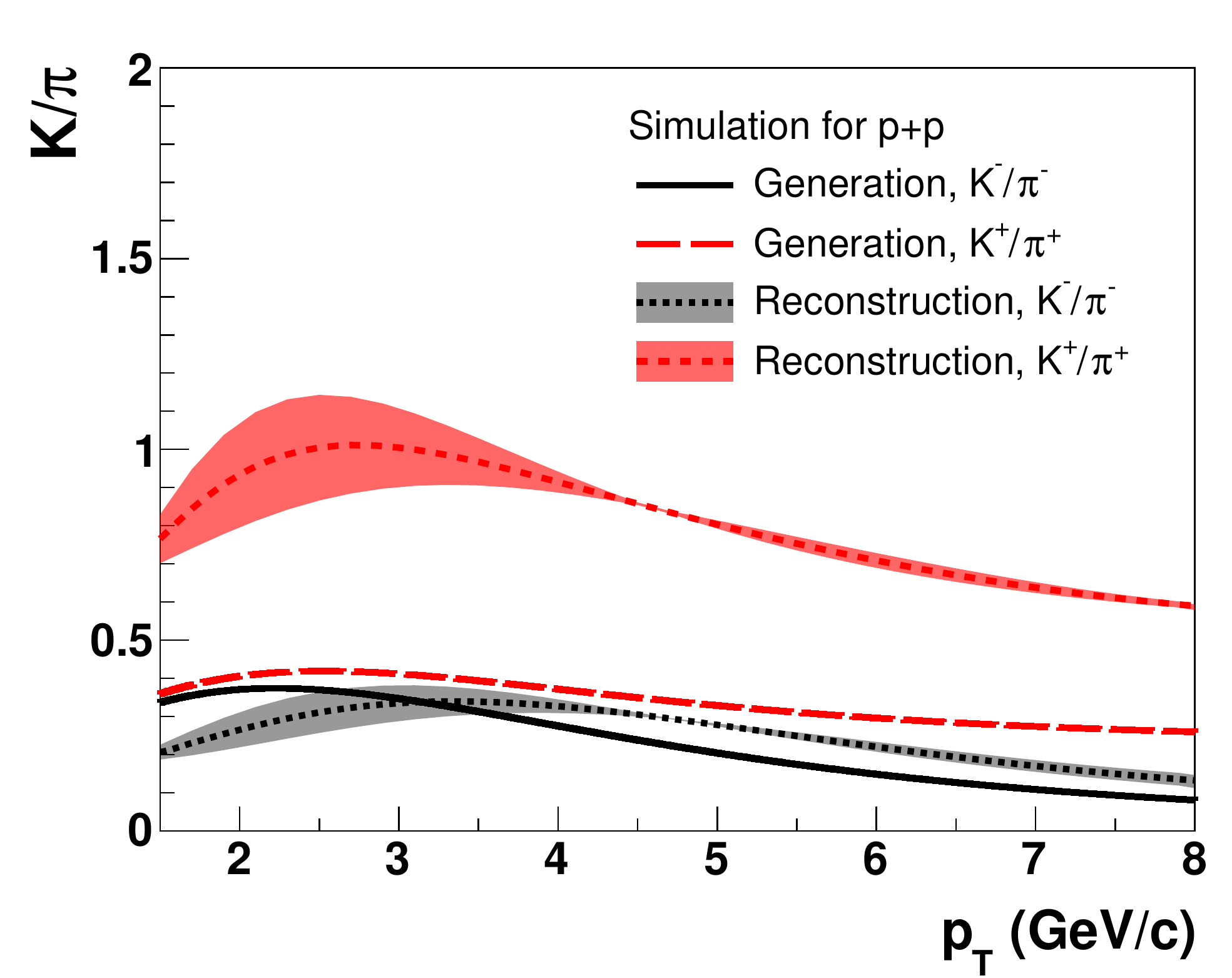}
\caption{\label{fig:Kpi_ratio_pp_pt}
Estimation of $K/\pi$ ratios at the collision vertex (Generation) and 
$K/\pi$ ratios for reconstructed muon arm tracks in in the {\sc geant4} 
simulation (Reconstruction) for \pp collisions. The variations from the 
different physics list in the {\sc geant4} simulation are shown as the 
bands on the reconstructed $K/\pi$ ratios. According to the simulation, 
the reconstructed particle composition is modified compared to the 
generated one due to interaction with the detector material.}
\end{figure}
\begin{figure}[thb]
\centering
\includegraphics[width=1.0\linewidth]{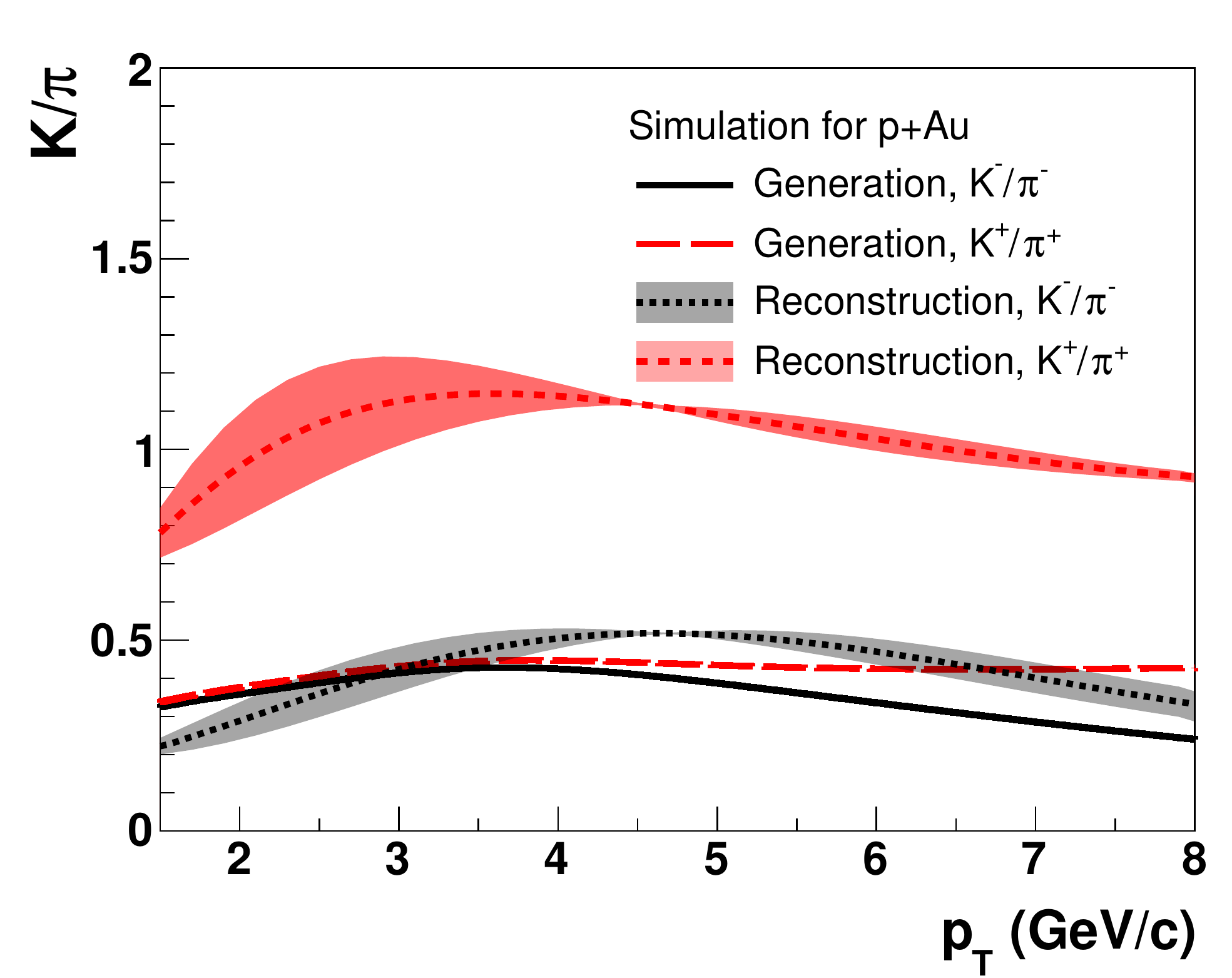}
\caption{\label{fig:Kpi_ratio_pAu_pt}
Estimation of $K/\pi$ ratios at the collision vertex (Generation) and 
$K/\pi$ ratios for reconstructed muon arm tracks in in the {\sc geant4} 
simulation (Reconstruction) for \pA collisions. The variations from the 
different physics lists in the {\sc geant4} simulation are shown as the 
bands on the reconstructed $K/\pi$ ratios. According to the simulation, 
the reconstructed particle composition is modified compared to the 
generated one due to interaction with the detector material.}
\end{figure}

An estimation of the particle composition in the measured charged hadron 
sample was developed in ~\cite{PHENIX:2012itj,PHENIX:2013txu}. Based on 
{\sc pythia}~\cite{Sjostrand:2006za} and {\sc 
hijing}~\cite{Gyulassy:1994ew} event generators. Charged hadron spectra 
measured at midrapidity in \pp and \dAu at 
RHIC~\cite{PHENIX:2011rvu,STAR:2011iap,PHENIX:2013kod} were extrapolated 
to the PHENIX muon arm rapidity ($1.2<\eta<2.4$) for \pp, \pAl, and \pAu 
collisions.  According to a {\sc geant4}~\cite{GEANT4:2002zbu} detector 
simulation (release 10.0.p02), the initial particle composition is 
modified due to interaction with the detector material including the 
front absorber. The reconstructed charged hadrons 
are mostly $K^{\pm}$ 
and $\pi^{\pm}$ where others ($p$, $\bar{p}$) are less than 10\%. The 
estimated $K/\pi$ ratios at the collision vertex and reconstructed 
ratios from the simulation are shown in Fig.~\ref{fig:Kpi_ratio_pp_pt} 
for \pp collisions and Fig.~\ref{fig:Kpi_ratio_pAu_pt} for \pA collisions. 
In both \pp and \pA collisions, the reconstructed $K^{-}$ are more suppressed than 
$K^{+}$ by the front absorber material due to its larger cross section 
with protons~\cite{Workman:2022ynf}. Momentum smearing also changes 
the shape of the ratios. 
Among the {\sc geant4} 
physics lists~\cite{Allison:2016lfl}, the default used is the QGSP-BERT,
which to simulate hadronic interactions applies the quark-gluon 
string model for high-energy and the Bertini cascade model for 
low-energy hadrons. To estimate possible variations, the FTFP-BERT and 
QGSP-BIC physics lists are used where the FTF model uses the FRITIOF 
description of string excitation and fragmentation and BIC uses the 
binary cascade model. The variations from the different physics lists in 
the {\sc geant4} simulation are shown in Figs. ~\ref{fig:Kpi_ratio_pp_pt} 
and~\ref{fig:Kpi_ratio_pAu_pt} as bands on the reconstructed $K/\pi$ ratios.

\subsection{Determination of the TSSA}

The unbinned maximum-likelihood method was used to extract the TSSA 
($A_N$) in this study. It was established in the previous study which 
used the same 
detectors~\cite{PHENIX:2017wbv,PHENIX:2018qvl,PHENIX:2019ouo}. The 
method is robust even for low-statistics data compared to binned 
approaches. The extended log-likelihood $\log\mathcal{L}$ is defined as

\begin{equation}
\label{eq:maxlikelihood1}
\log \mathcal{L} = \sum_i \log(1 + P \cdot A_N \sin(\phi_{\rm pol} - \phi_h^i)) + \textrm{constant},
\end{equation}

\noindent where $P$ is the proton beam polarization and $\phi_h^i$ is 
the azimuthal angle of the $i$-th hadron with respect to the incoming 
polarized-proton-beam direction. The $\phi_{\rm pol}$ is the azimuthal 
angle for the beam polarization direction, which takes the values of 
$+\frac{\pi}{2}/-\frac{\pi}{2}$ for $\uparrow/\downarrow$ polarized beam 
bunches in 2015. The $A_N$ is chosen to be the value which maximizes 
$\log \mathcal{L}$. The statistical uncertainty of the log-likelihood 
estimator is calculated from the second derivative of $\mathcal{L}$ with 
respect to $A_N$,

\begin{equation}
\label{eq:maxlikelihood_err1}
\sigma^2(A_N) = \left(-\frac{\partial^2 \mathcal{L}}{\partial A_N^2}\right)^{-1}.
\end{equation}

In $p$$+$$p$ collisions, both beams were polarized. The $A_N$s were measured 
separately for each beam and turned out to be consistent with each 
other. They were averaged for the final asymmetry.

The $A_N$ is checked with the azimuthal-fitting method used in previous 
analyses~\cite{PHENIX:2014qwb,PHENIX:2017wbv,PHENIX:2018qvl,PHENIX:2019ouo} 
based on the polarization formula~\cite{Ohlsen:1973wf}:
\begin{equation}
\label{eq:anvsphi_pp}
A_N(\phi)=\frac{\sigma^{\uparrow}(\phi)-\sigma^{\downarrow }(\phi)}
{\sigma^{\uparrow}(\phi)+\sigma^{\downarrow}(\phi)}\\
=\frac{1}{P}\cdot\frac{N^{\uparrow}(\phi)
-R\cdot N^{\downarrow}(\phi)}{N^{\uparrow}(\phi)
+R\cdot N^{\downarrow}(\phi)},\\
\end{equation}
where $A_N(\phi)$ indicates the simple-count-based transverse 
single-spin asymmetry calculated for each of 16 azimuthal $\phi$-bins; 
$\sigma^{\uparrow }$, $\sigma^{\downarrow}$ represent cross sections; 
and $N^{\uparrow}$, $N^{\downarrow}$ are yields for each polarization of 
spin up ($\uparrow$) or down ($\downarrow$). $R = 
L^{\uparrow}/L^{\downarrow}$ is the ratio of luminosities (relative 
luminosity) between bunches with spin up and spin down where the luminosity 
is determined by sampled counts from minimum-bias triggers for the 
corresponding spin orientation. The $A_N$ in this method is calculated from 
the fit of $A_N(\phi)$ distributions using a function $A_N\cdot 
\sin(\phi_{\rm pol\uparrow} - \phi)$ where $\phi_{\rm pol 
\uparrow}=\pi/2$ indicates the azimuthal direction of upward polarized 
bunches. The difference of the $A_N$'s for this method and the 
maximum-likelihood method is conservatively included in the systematic 
uncertainty.

The interactions between particles and material prior to being detected 
by the MuTr, and the finite resolution for momentum and azimuthal angle 
$\phi$, may result in a kinematic smearing effect for the $A_N$ results. 
The full {\sc geant4} simulation is used to correct this effect. The 
effect from $\phi$ smearing was found to be negligible.

\begin{figure}[thb]
\centering
\includegraphics[width=1.0\linewidth]{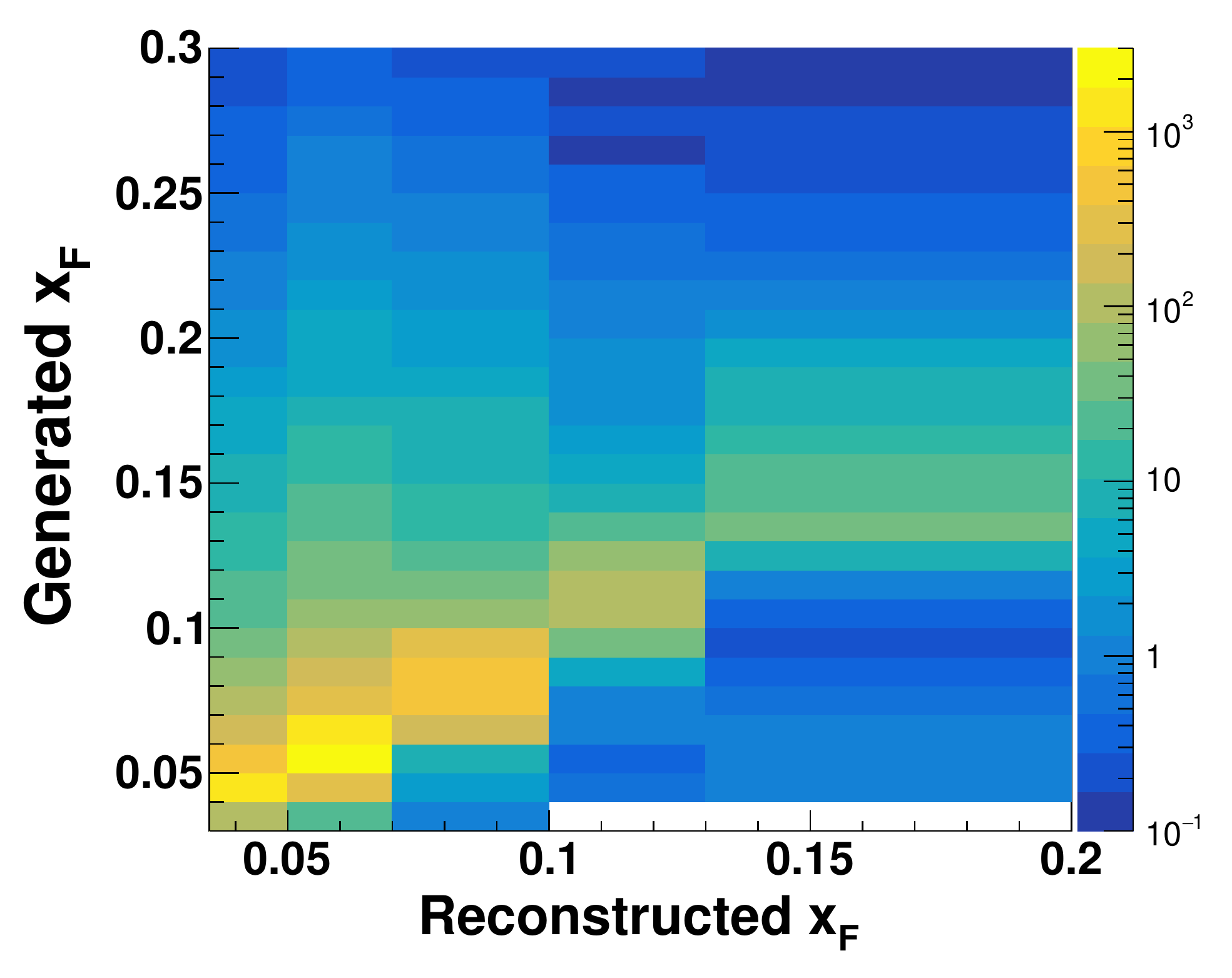}
\caption{\label{fig:unfolding_2D_xF}
The distribution of generated $x_F$ for reconstructed $x_F$ bins 
from {\sc geant4} simulation in \pp collisions.}
\end{figure}

The momentum smearing is represented in Fig.~\ref{fig:unfolding_2D_xF}, 
where the generated $x_F$ distribution for reconstructed $x_F$ bin 
is shown. The momentum 
smearing effect is corrected by resolving a set of linear equations 
which connect $A_N$ for reconstructed $x_F$ ($p_T$) bins 
($A_N^{\rm reco}$) and $A_N$ for true (generated) $x_F$ ($p_T$) bins 
($A_N^{\rm true}$):
\begin{equation}
\label{eq:unfold1}
A_N^{{\rm reco}, m} =  \sum_{i} f^{i \rightarrow m} \cdot A_N^{{\rm true}, i},
\end{equation}
where $A_N^{{\rm true}, i}$ represents $A_N$ for the $i$-th true 
momentum ($x_F$, $p_T$) bin and $A_N^{{\rm reco}, m}$ is $A_N$ for the 
reconstructed momentum in the $m$-th momentum bin. $f^{i \rightarrow m}$ 
is the fraction of charged hadrons reconstructed in the $m$-th momentum 
bin from the $i$-th true (generated) momentum bin in the simulation. 
$A_N^{\rm reco}$ is measured including an underflow and overflow bin for 
$0.035<x_F<0.3$ and $1.25<p_T<15{\rm\ GeV}/c$, and then $A_N^{\rm true}$ 
is calculated for $0.04<x_F<0.2$ and $1.5<p_T<7{\rm\ GeV}/c$. The 
difference between $A_N^{\rm reco}$ and $A_N^{\rm true}$ was found to be 
small, and was conservatively included in systematic uncertainties. The 
variation of $A_N$ due to muon contribution is minimal ($<0.0005$) and 
not included in the systematic uncertainty.

\section{Results} 
\label{sec:results}

\subsection{Results in Proton-Proton Collisions}

\begin{figure}[thb]
\centering
\includegraphics[width=1.0\linewidth]{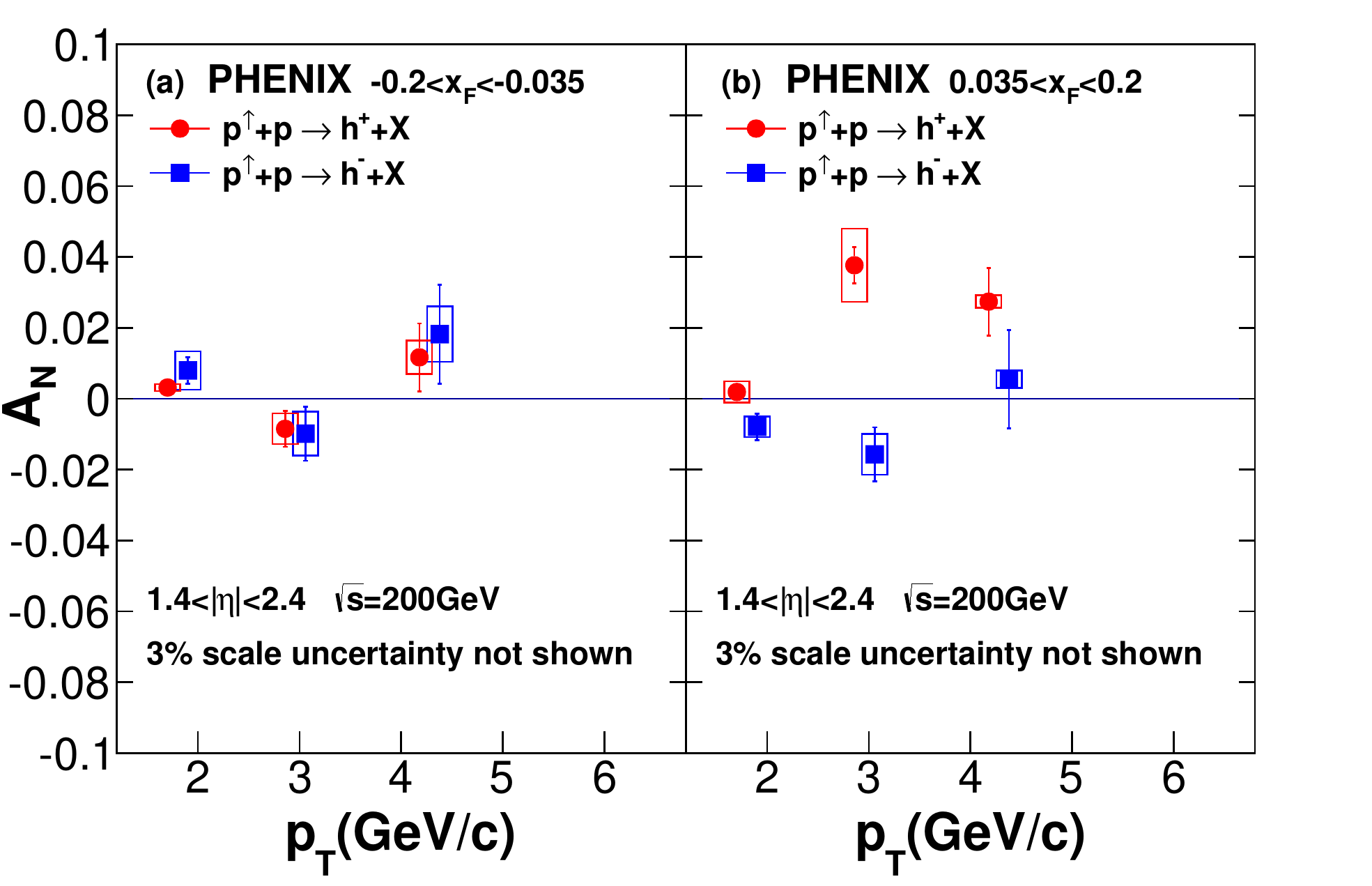}
\caption{\label{fig:AN_pt_pp}
$A_N$ of charged hadrons from \polpp collisions as a function of $p_T$ 
in the (a) backward ($x_F<0$,) and (b) forward ($x_F>0$) regions. 
Vertical bars (boxes) represent statistical (systematic) uncertainties. 
Points are shifted by $p_T=$+0.2 GeV/$c$ for negatively charged 
hadrons. A scale uncertainty from the polarization (3\%) is not 
included.}
\end{figure}

\begin{figure}[thb]
\centering
\includegraphics[width=1.0\linewidth]{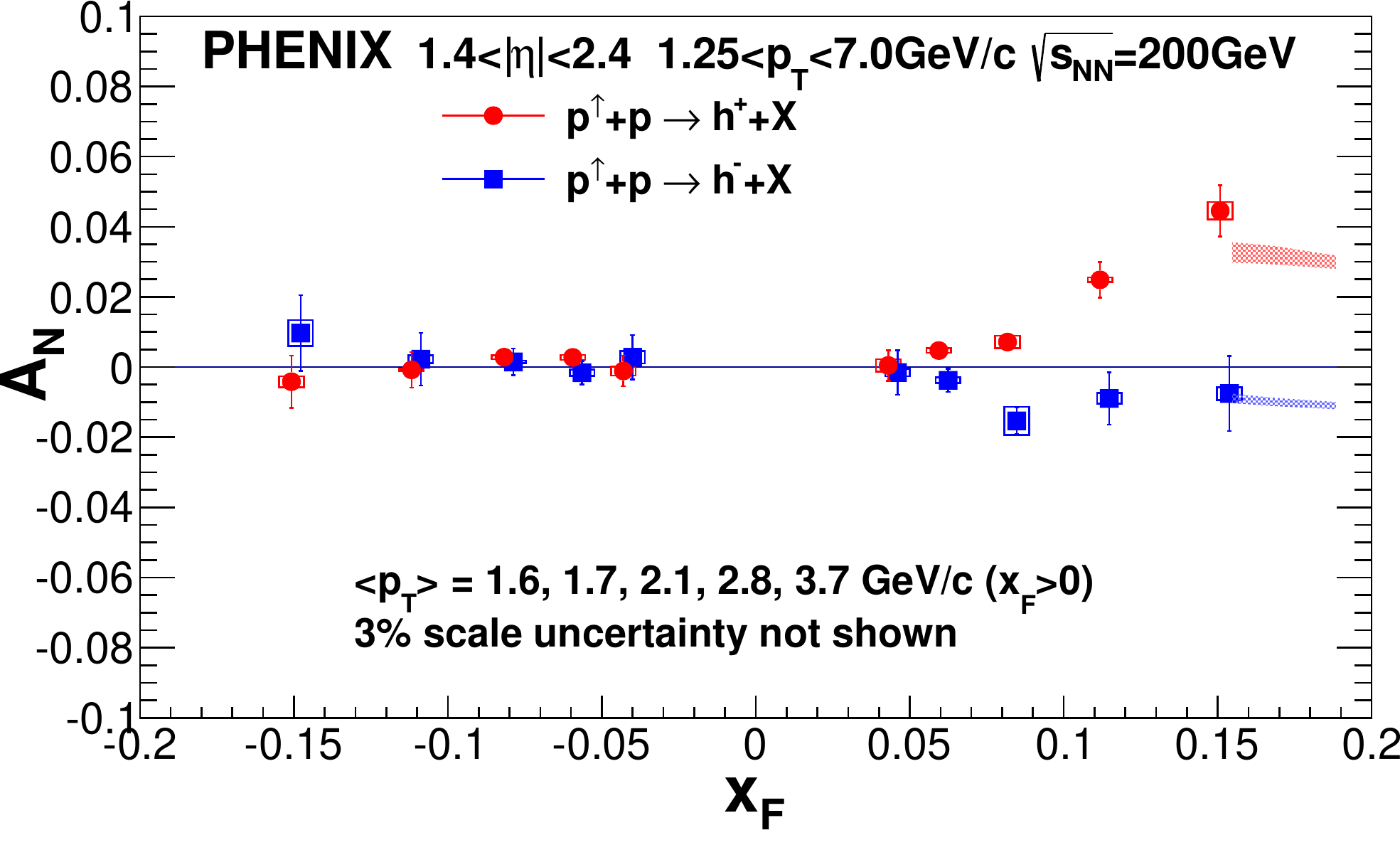}
\caption{\label{fig:AN_xf_pp_band}
$A_N$ of charged hadrons from \polpp collisions as a function of $x_F$, 
where $x_F>0$ is along the direction of the polarized proton. Vertical 
bars (boxes) represent statistical (systematic) uncertainties. Points 
are shifted by $x_F=$+0.003 for negatively charged hadrons. Dotted bands 
represent twist-3 model calculations~\cite{Gamberg:2017gle} by varying 
the $K/\pi$ ratio by $\pm 30 \%$ of the central value. A scale uncertainty
from the polarization (3\%) is not included.}
\end{figure}

The TSSAs in the production of charged hadrons at \absetarange in 
transversely-polarized proton-proton collisions (\polpp) are shown in 
Fig.~\ref{fig:AN_pt_pp} as a function of $p_T$ and in 
Fig.~\ref{fig:AN_xf_pp_band} as a function of $x_F$. The results are listed in 
Tables~\ref{tab:AN_pt_neg_pp},~\ref{tab:AN_pt_pos_pp},~\ref{tab:AN_xf_neg_pp}, 
and~\ref{tab:AN_xf_pos_pp}. In the tables, $\delta A_N^{\rm stat}$ is 
the statistical uncertainty, and $\delta A_N^{\rm syst}$ is the total 
systematic uncertainty. The systematic uncertainty is obtained from the 
quadratic sum of two components ($\delta A_N^{\rm method}$ and $\delta 
A_N^{\rm smear}$). As explained in the previous section, $\delta 
A_N^{\rm method}$ is the difference between the two methods of 
determining $A_N$, while $\delta A_N^{\rm smear}$ is the difference 
caused by the momentum migration correction (Eq.~\ref{eq:unfold1}). 
Three different physics lists are tested in the correction and the 
variation among them is negligible.

On the panel (a) of Fig.~\ref{fig:AN_pt_pp}, $A_N$ at backward rapidity 
($x_F<0$) for charged hadrons is consistent with zero within 
uncertainty. On the panel (b), $A_N$ at forward rapidity ($x_F>0$) is 
positive for positively charged hadrons. Positive asymmetry is also 
shown in the $x_F$-binned result in Fig.~\ref{fig:AN_xf_pp_band}. $A_N$ 
for positively charged hadrons at $x_F>0$ shows an increasing trend with 
respect to $x_F$. The result for negatively charged hadrons shows some 
indication of negative $A_N$ around $x_F>0.07$.

The previous results at RHIC energies with larger $|x_F|$ and $|\eta|$ 
than this measurement showed positive $A_N$ for $\pi^{+}$, $K^{\pm}$ and 
negative $A_N$ for $\pi^{-}$ at $x_F{}>0$~\cite{BRAHMS:2008doi,Lee:2007zzh}. 
In this measurement, the result for $h^{+}$ is 
comparable with $A_N$ for $\pi^{+}$, $K^{+}$. $A_N$ for $h^{+}$ is 
increasing as a function of $x_F$ at $x_F>0$ and is consistent with zero 
at $x_F<0$. $\pi^{-}$ and $K^{-}$ in the previous measurements showed 
the opposite sign of $A_N$ at forward rapidity. Therefore one may expect 
the $A_N$ for $h^{-}$ has been partially canceled and this can explain 
the smaller size of $A_N$ for $h^{-}$ at $x_F>0$ in 
Fig.~\ref{fig:AN_xf_pp_band}. This is also confirmed in twist-3 model 
calculations shown in Fig.~\ref{fig:AN_xf_pp_band}. The dotted bands in 
Fig.~\ref{fig:AN_xf_pp_band} are obtained from $K/\pi$ ratios in the 
simulation combined with theoretical $A_N$ for charged pions and kaons 
from twist-3 calculation for $x_F>0.15$~\cite{Gamberg:2017gle}, which 
describes previous RHIC results at forward 
rapidity~\cite{BRAHMS:2008doi,Lee:2007zzh}. In addition, the effect of 
the $K/\pi$ production ratio variation is tested for a relative $\pm 
30\%$ variation and the width of the band shows the possible variation 
of $A_N$ mixture for $h^{+}$ and $h^{-}$. The variation of the $K/\pi$ 
production ratio from different {\sc geant4} physics lists shown in 
Fig.~\ref{fig:Kpi_ratio_pp_pt} is much smaller than the relative $\pm 
30\%$ that was considered.

\subsection{Results in Proton-Nucleus Collisions}

\begin{figure}[thb]
\centering
\includegraphics[width=1.0\linewidth]{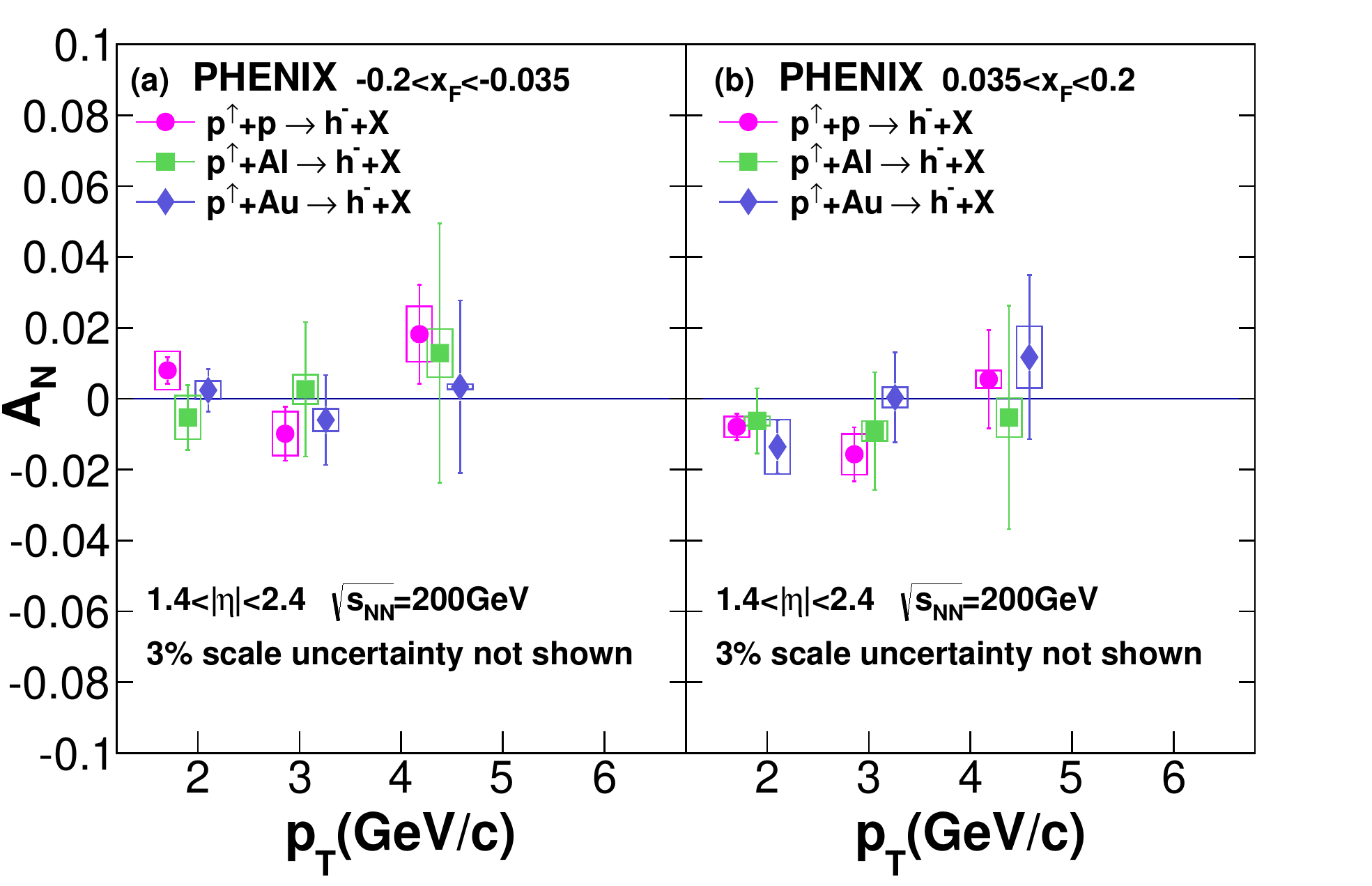}
\caption{\label{fig:AN_pt_neg}
$A_N$ of negatively charged hadrons from \polpp, \polpAl, and \polpAu 
collisions as a function of $p_T$ in the (a) backward ($x_F<0$) and (b) 
forward ($x_F>0$) regions. Vertical bars (boxes) represent statistical 
(systematic) uncertainties. Points are shifted by $p_T=$+0.02 (+0.04) 
GeV/$c$ for \polpAl (\polpAu) results. A scale uncertainty from the 
polarization (3\%) is not included.}
\end{figure}

\begin{figure}[thb]
\includegraphics[width=1.0\linewidth]{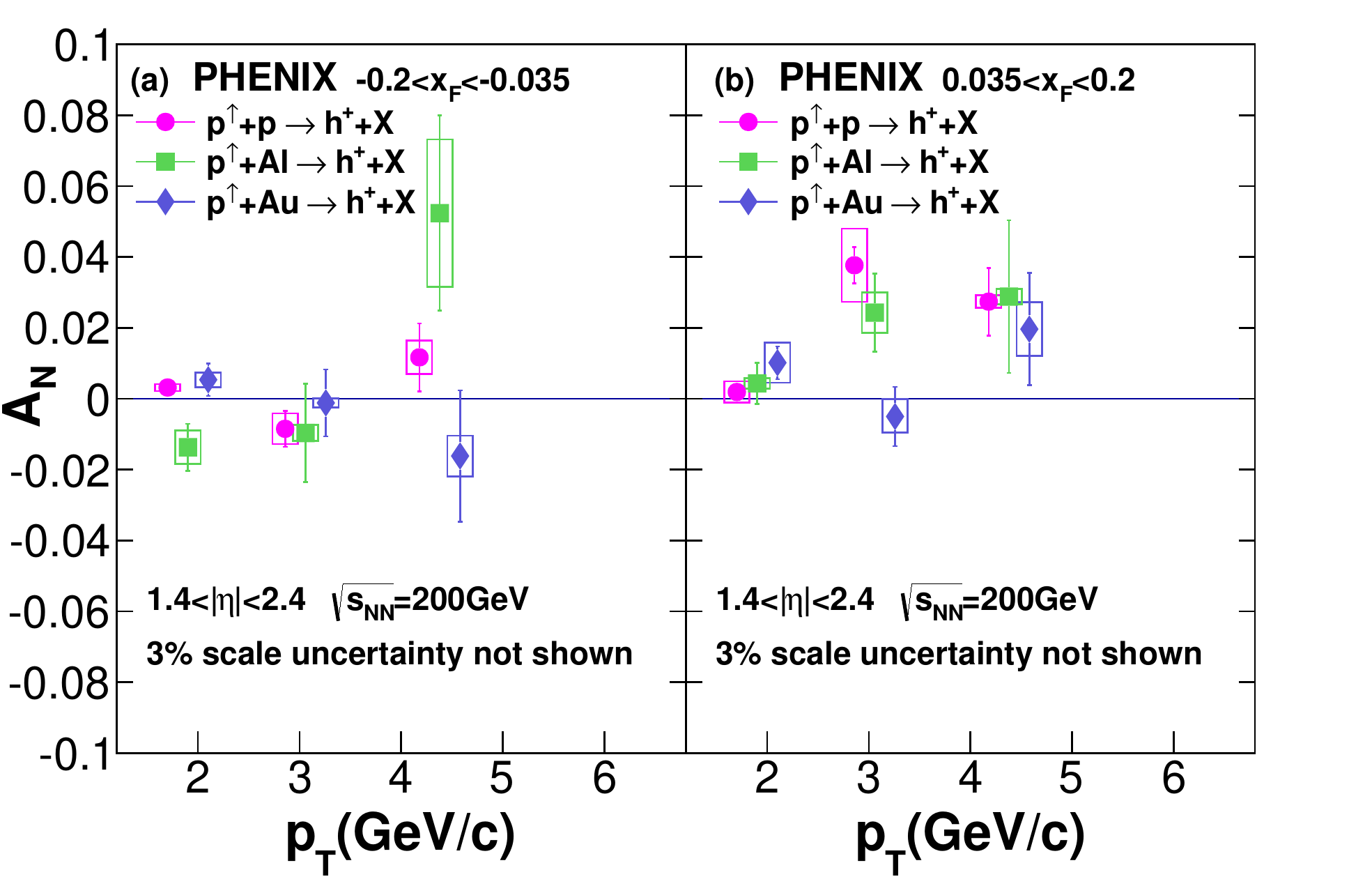}
\caption{\label{fig:AN_pt_pos}
$A_N$ of positively charged hadrons from \polpp, \polpAl, and \polpAu 
collisions as a function of $p_T$ in the (a) backward ($x_F<0$) and (b) 
forward ($x_F>0$) regions. Vertical bars (boxes) represent statistical 
(systematic) uncertainties. Points are shifted by $p_T=$+0.02 (+0.04) 
GeV/$c$ for \polpAl (\polpAu) results. A scale uncertainty from the 
polarization (3\%) is not included.}
\end{figure}

\begin{figure}[thb]
\includegraphics[width=1.0\linewidth]{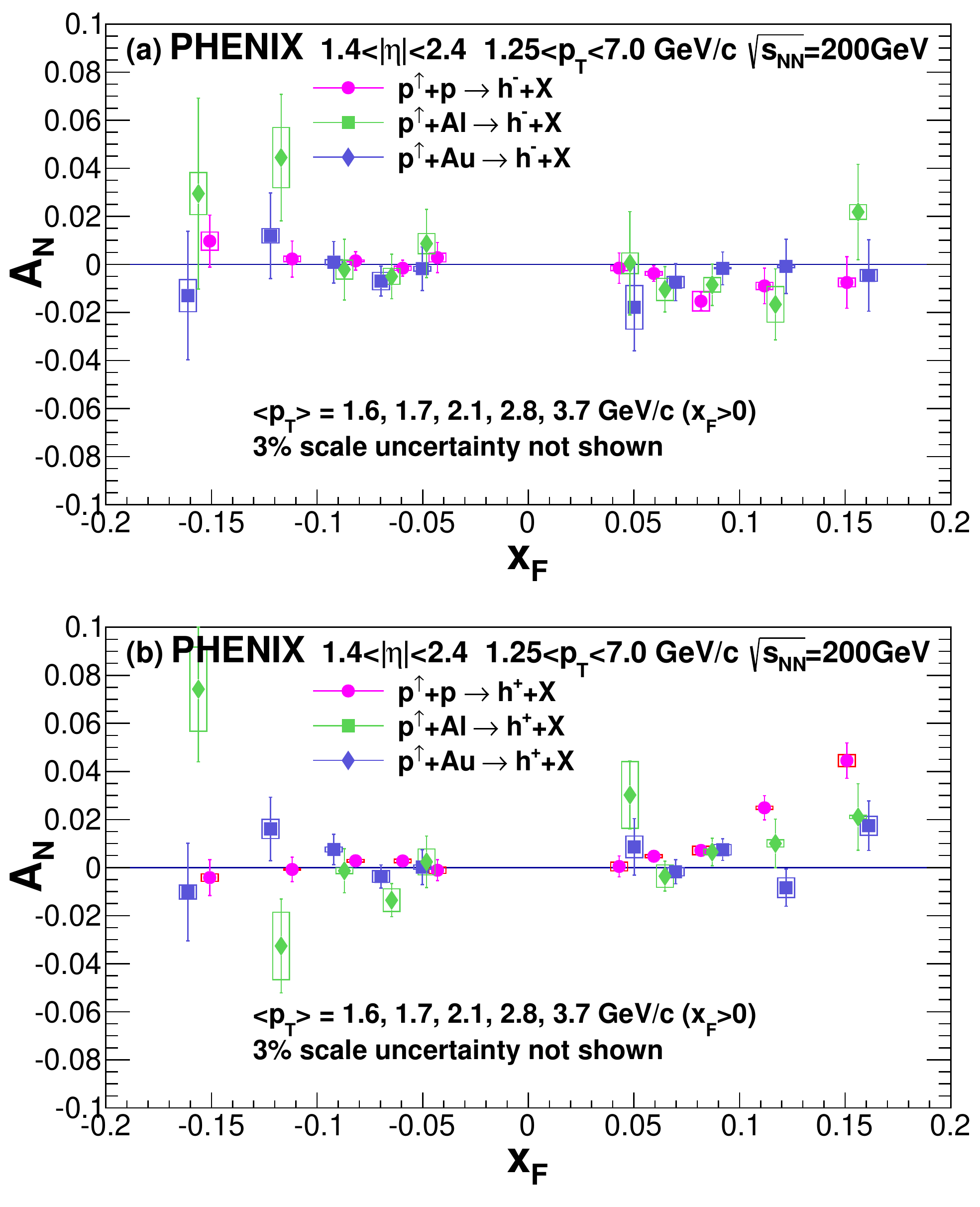}
\caption{\label{fig:AN_xf_bothchg}
$A_N$ of (a) negatively charged hadrons and (b) positively charged 
hadrons from \polpp, \polpAl, and \polpAu collisions as a function of 
$x_F$, where $x_F>0$ is along the direction of the polarized proton. 
Vertical bars (boxes) represent statistical (systematic) uncertainties. 
Points are shifted by $x_{F}= -0.01, -0.005, +0.005, +0.01\ (-0.005, 
-0.003, +0.003, +0.005)$ at $-0.2<x_{F}<-0.05, -0.05<x_{F}<0.04, 
0.04<x_{F}<0.05, 0.05<x_{F}<0.2$, respectively for \polpAu (\polpAl) 
results. A scale uncertainty from the polarization (3\%) is not 
included.}
\end{figure}

The $A_N$ results of charged hadrons in proton-nucleus collisions 
(\polpAl and \polpAu) are shown in 
Fig.~\ref{fig:AN_pt_neg},~\ref{fig:AN_pt_pos}, 
and~\ref{fig:AN_xf_bothchg} with \polpp results and listed in 
Tables~\ref{tab:AN_pt_neg_pAl},~\ref{tab:AN_pt_pos_pAl},
~\ref{tab:AN_xf_neg_pAl}, 
and~\ref{tab:AN_xf_pos_pAl} for \polpAl and 
\ref{tab:AN_pt_neg_pAu},~\ref{tab:AN_pt_pos_pAu},~\ref{tab:AN_xf_neg_pAu}, 
and~\ref{tab:AN_xf_pos_pAu} for \polpAu. Results at $x_F<0$ in all 
collision systems (\polpp, \polpA) are close to zero; the 
combined $A_N$ are within $1.5\sigma$ from zero where 
$\sigma$ is total uncertainty.
Results for negatively charged hadrons from \polpAl and 
\polpAu collisions shown in Fig.~\ref{fig:AN_pt_neg} and 
\ref{fig:AN_xf_bothchg} are also close to zero in terms of the total 
uncertainty; the combined $A_N$ are within $1.2\sigma$ from zero. 
In contrast, $A_N$ in \polpp collisions shows some indication of 
a negative asymmetry at $p_T<3.5 \rm{\ GeV}/c$ on panel (b) 
($x_{F}>0$) of Fig.~\ref{fig:AN_pt_neg} and at $0.07<x_{F}<0.10$ on 
panel (a) of Fig.~\ref{fig:AN_xf_bothchg}. However, the difference of 
$A_{N}$ between \polpAu and \polpp is not significant to state any 
modification of $|A_N|$ in \polpAu compared to \polpp results.

The $A_N$ for positively charged hadrons at $x_F>0$ in $\polpAu$ 
collisions is consistent with zero while the results for \polpp 
represents a positive asymmetry on panel (b) in Fig.~\ref{fig:AN_pt_pos} 
and on panel (b) in Fig.~\ref{fig:AN_xf_bothchg}.  The 
\polpAl results for $h^{+}$ at $x_F>0$ show some indication of a 
positive asymmetry, which is smaller than in \polpp collisions.

\section{Discussion}
\label{sec:discussion}

The results for the production of positively charged hadrons ($h^{+}$) 
in \polpp collisions agree with a trend with other RHIC data where 
$A_{N}$ for $\pi^{+}$ and $K^{+}$ is positive and increasing with 
respect to $x_{F}$ in the forward region 
($x_F>0$)~\cite{Lee:2007zzh,BRAHMS:2008doi}. The results for $A_N$ for 
negatively charged hadrons ($h^{-}$) range from slightly negative to 
zero at $x_F>0$; this may be caused by cancellation between positive 
asymmetry for $K^{-}$ and negative asymmetry for $\pi^{-}$ shown in RHIC 
data at larger $x_{F}$~\cite{Lee:2007zzh,BRAHMS:2008doi}. These agree 
with the trend of the theoretical calculation at 
$x_F>0.15$~\cite{Gamberg:2017gle}. In all collision systems, $A_{N}$ at 
backward rapidity ($x_F<0$) is close to zero asymmetry in terms of 
the total uncertainty, which also has been shown in previous RHIC 
measurements in \polpp collisions 
~\cite{STAR:2008ixi,BRAHMS:2008doi,PHENIX:2013wle}. 
The $A_{N}$ of positively charged hadrons at $0.1<x_{F}<0.2$ in \polpAl
and \polpAu is smaller than in \polpp. However, the total uncertainty 
of \polpAl results is too large to provide clear separation from \polpp 
or \polpAu results while \polpAu shows nuclear dependence.


Recent calculations using collinear factorization on polarized proton 
($p^{\uparrow}$) and the CGC framework on nuclear target 
($A$)~\cite{Schafer:2014zea} predicted that the $A_N$ of inclusive 
hadrons at forward rapidity in \polpA collisions has two contributions 
where one is $A$-independent and the other is 
$A^{-1/3}$-dependent~\cite{Hatta:2016wjz,Hatta:2016khv,Zhou:2017sdx,Benic:2018amn}. 
However, the $A^{-1/3}$-dependent term is dominant for $p_{T}$ less than 
$Q_{s}$ and $A$-independent term is dominant for higher $p_T$. In the 
kinematics of this measurement according to a recent 
calculation~\cite{Benic:2018amn}, $Q_{s}^{\rm Au}\approx 0.9$ GeV whereas 
our measurements correspond to $\langle p_{T}\rangle \approx 2.9 {\rm\ 
GeV}/c$ at $0.1<x_{F}<0.2$. An approach using lensing mechanism predicts 
that $A_{N}$ decreases as the atomic number of the target ($A$) 
increases for $k_{T}$ below or near $Q_{s}$~\cite{Kovchegov:2020kxg}.  A 
recent STAR result of $\pi^{0}$ in the more forward region at 
$2.6<\eta<4.0$, $p_T>1.5~{\rm GeV}/c$, and $0.2<x_{F}<0.7$ in \polpp and 
\polpAu collisions shows a smaller $A$-dependence~\cite{STAR:2020grs} 
than what is observed 
here. Given that our measurement and that from STAR 
used different kinematic ranges and hadron species, further detailed 
studies of various observables within a wide kinematic range will be 
informative on the origin of $A_N$ and its interplay with small-$x$ 
phenomena.

\section{Summary}
\label{sec:summary}

Reported here are the transverse single-spin asymmetry ($A_N$) of 
positively and negatively charged hadrons ($h^{\pm}$) in forward and 
backward rapidity ($1.4<|\eta|<2.4$) over the range of $1.5<p_{T}<7.0 
{\rm\ GeV}/c$ and $0.04<|x_{F}|<0.2$ from transversely polarized 
proton-proton (\polpp) and proton-nucleus (\polpA) collisions. The 
results at $x_{F}<0$ are close to zero at all systems. In 
$x_{F}>0$, negative charged hadron results show small to zero $A_{N}$ in 
\polpp collisions and nearly zero $A_N$ in \polpA collisions. $A_{N}$ for 
positively charged hadrons increases to positive values as $x_{F}$ 
increases in \polpp collisions, but \polpAu results show suppression of 
$A_{N}$ at $0.1<x_{F}<0.2$ compared to the \polpp result. The results 
will aid in understanding of the origin of $A_{N}$ and offer a tool to 
investigate nuclear effects and phenomena in small-$x$.



\begin{acknowledgments}

We thank the staff of the Collider-Accelerator and Physics
Departments at Brookhaven National Laboratory and the staff of
the other PHENIX participating institutions for their vital
contributions.  
We also thank D. Pitonyak for the theory calculation. 
We acknowledge support from the Office of Nuclear Physics in the
Office of Science of the Department of Energy,
the National Science Foundation,
Abilene Christian University Research Council,
Research Foundation of SUNY, and
Dean of the College of Arts and Sciences, Vanderbilt University
(U.S.A),
Ministry of Education, Culture, Sports, Science, and Technology
and the Japan Society for the Promotion of Science (Japan),
Natural Science Foundation of China (People's Republic of China),
Croatian Science Foundation and
Ministry of Science and Education (Croatia),
Ministry of Education, Youth and Sports (Czech Republic),
Centre National de la Recherche Scientifique, Commissariat
{\`a} l'{\'E}nergie Atomique, and Institut National de Physique
Nucl{\'e}aire et de Physique des Particules (France),
J. Bolyai Research Scholarship, EFOP, the New National Excellence
Program ({\'U}NKP), NKFIH, and OTKA (Hungary),
Department of Atomic Energy and Department of Science and Technology
(India),
Israel Science Foundation (Israel),
Basic Science Research and SRC(CENuM) Programs through NRF
funded by the Ministry of Education and the Ministry of
Science and ICT (Korea).
Ministry of Education and Science, Russian Academy of Sciences,
Federal Agency of Atomic Energy (Russia),
VR and Wallenberg Foundation (Sweden),
University of Zambia, the Government of the Republic of Zambia (Zambia),
the U.S. Civilian Research and Development Foundation for the
Independent States of the Former Soviet Union,
the Hungarian American Enterprise Scholarship Fund,
the US-Hungarian Fulbright Foundation,
and the US-Israel Binational Science Foundation.

\end{acknowledgments}


\appendix*

\section{DATA TABLES}
\label{appendix}

Tables~\ref{tab:AN_pt_neg_pp},~\ref{tab:AN_pt_pos_pp},~\ref{tab:AN_xf_neg_pp}, 
and~\ref{tab:AN_xf_pos_pp} list the TSSAs in the production of charged 
hadrons at \absetarange in transversely-polarized proton-proton 
collisions (\polpp) that are shown in Fig.~\ref{fig:AN_pt_pp} as a 
function of $p_T$ and in Fig.~\ref{fig:AN_xf_pp_band} as a function of 
$x_F$. See further details in the subsection ``A. Results in Proton-Nucleus 
Collisions."

Tables~\ref{tab:AN_pt_neg_pAl},~\ref{tab:AN_pt_pos_pAl}, 
~\ref{tab:AN_xf_neg_pAl}, and~\ref{tab:AN_xf_pos_pAl} for \polpAl and 
\ref{tab:AN_pt_neg_pAu},~\ref{tab:AN_pt_pos_pAu},~\ref{tab:AN_xf_neg_pAu}, 
and~\ref{tab:AN_xf_pos_pAu} for \polpAu. list the $A_N$ results of 
charged hadrons in proton-nucleus collisions (\polpAl and \polpAu) that 
are shown in Figs.~\ref{fig:AN_pt_neg},~\ref{fig:AN_pt_pos}, 
and~\ref{fig:AN_xf_bothchg} with \polpp results.
See further details in the subsection 
``B. Results in Proton-Nucleus Collisions."

\begingroup \squeezetable
\begin{table*}[tbh]
\begin{minipage}{0.48\linewidth}
\caption{\label{tab:AN_pt_neg_pp}
Data table of $A_{N}$ for negatively charged hadron ($h^{-}$) in 
transversely-polarized \pp collisions as a function of $p_T$.}
\begin{ruledtabular}
\begin{tabular}{cccccc}
$p_{T}$(GeV/$c$) & $A_{N}$ & $\delta A_N^{\rm stat}$ & $\delta A_N^{\rm syst}$ & $\delta A_N^{\rm method}$ & $\delta A_N^{\rm smear}$ \\ \hline 
\multicolumn{6}{c}{Forward ($x_F>0$)} \\
(1.50, 2.50) & $-0.008$ & $\pm$0.004 & $\pm0.003$ & $\pm0.001$  & $\pm0.003$\\
(2.50, 3.50) & $-0.016$ & $\pm$0.008 & $\pm0.006$ & $\pm0.001$  & $\pm0.006$\\
(3.50, 7.00) & $0.005$ & $\pm$0.014 & $\pm0.002$ & $\pm0.002$  & $\pm0.002$\\
\multicolumn{6}{c}{Backward ($x_F<0$)} \\
(1.50, 2.50) & $0.008$ & $\pm$0.004 & $\pm0.005$ & $\pm0.000$  & $\pm0.005$\\
(2.50, 3.50) & $-0.010$ & $\pm$0.008 & $\pm0.006$ & $\pm0.001$  & $\pm0.006$\\
(3.50, 7.00) & $0.018$ & $\pm$0.014 & $\pm0.008$ & $\pm0.001$  & $\pm0.008$\\
\end{tabular}
\end{ruledtabular}
\end{minipage}
\hspace{0.2cm}
\begin{minipage}{0.48\linewidth}
\caption{\label{tab:AN_pt_pos_pp}
Data table of $A_{N}$ for positively charged hadron ($h^{+}$) in 
transversely-polarized \pp collisions as a function of $p_T$.}
\begin{ruledtabular}
\begin{tabular}{cccccc}
$p_{T}$(GeV/$c$) & $A_{N}$ & $\delta A_N^{\rm stat}$ & $\delta A_N^{\rm syst}$ & $\delta A_N^{\rm method}$ & $\delta A_N^{\rm smear}$ \\ \hline 
\multicolumn{6}{c}{Forward ($x_F>0$)} \\
(1.50, 2.50) & $0.002$ & $\pm$0.002 & $\pm0.003$ & $\pm0.000$  & $\pm0.003$\\
(2.50, 3.50) & $0.038$ & $\pm$0.005 & $\pm0.010$ & $\pm0.001$  & $\pm0.010$\\
(3.50, 7.00) & $0.027$ & $\pm$0.010 & $\pm0.002$ & $\pm0.001$  & $\pm0.002$\\
\multicolumn{6}{c}{Backward ($x_F<0$)} \\
(1.50, 2.50) & $0.003$ & $\pm$0.002 & $\pm0.001$ & $\pm0.001$  & $\pm0.001$\\
(2.50, 3.50) & $-0.008$ & $\pm$0.005 & $\pm0.004$ & $\pm0.001$  & $\pm0.004$\\
(3.50, 7.00) & $0.012$ & $\pm$0.010 & $\pm0.005$ & $\pm0.001$  & $\pm0.004$\\
\end{tabular}
\end{ruledtabular}
\end{minipage}
\begin{minipage}{0.48\linewidth}
\caption{\label{tab:AN_xf_neg_pp}
Data table of $A_{N}$ for negatively charged hadron ($h^{-}$) in 
transversely-polarized \pp collisions as a function of $x_F$.}
\begin{ruledtabular}
\begin{tabular}{cccccc}
$x_{F}$ & $A_{N}$ & $\delta A_N^{\rm stat}$ & $\delta A_N^{\rm syst}$ & $\delta A_N^{\rm method}$ & $\delta A_N^{\rm smear}$ \\ \hline 
$(-0.200, -0.130)$ & $0.010$ & $\pm$0.011 & $\pm0.004$ & $\pm0.003$  & $\pm0.002$\\
$(-0.130, -0.100)$ & $0.002$ & $\pm$0.008 & $\pm0.001$ & $\pm0.001$  & $\pm0.000$\\
$(-0.100, -0.070)$ & $0.001$ & $\pm$0.004 & $\pm0.000$ & $\pm0.000$  & $\pm0.000$\\
$(-0.070, -0.050)$ & $-0.002$ & $\pm$0.003 & $\pm0.001$ & $\pm0.000$  & $\pm0.001$\\
$(-0.050, -0.040)$ & $0.003$ & $\pm$0.006 & $\pm0.002$ & $\pm0.001$  & $\pm0.002$\\
$(0.040, 0.050)$ & $-0.002$ & $\pm$0.006 & $\pm0.001$ & $\pm0.000$  & $\pm0.001$\\
$(0.050, 0.070)$ & $-0.004$ & $\pm$0.003 & $\pm0.001$ & $\pm0.000$  & $\pm0.001$\\
$(0.070, 0.100)$ & $-0.015$ & $\pm$0.004 & $\pm0.004$ & $\pm0.001$  & $\pm0.004$\\
$(0.100, 0.130)$ & $-0.009$ & $\pm$0.007 & $\pm0.002$ & $\pm0.000$  & $\pm0.001$\\
$(0.130, 0.200)$ & $-0.008$ & $\pm$0.011 & $\pm0.002$ & $\pm0.001$  & $\pm0.001$\\
\end{tabular}
\end{ruledtabular}
\end{minipage}
\hspace{0.2cm}
\begin{minipage}{0.48\linewidth}
\caption{\label{tab:AN_xf_pos_pp}
Data table of $A_{N}$ for positively charged hadron ($h^{+}$) in 
transversely-polarized \pp collisions as a function of $x_F$.}
\begin{ruledtabular}
\begin{tabular}{cccccc}
$x_{F}$ & $A_{N}$ & $\delta A_N^{\rm stat}$ & $\delta A_N^{\rm syst}$ & $\delta A_N^{\rm method}$ & $\delta A_N^{\rm smear}$ \\ \hline 
$(-0.200, -0.130)$ & $-0.004$ & $\pm$0.007 & $\pm0.002$ & $\pm0.002$  & $\pm0.000$\\
$(-0.130, -0.100)$ & $-0.001$ & $\pm$0.005 & $\pm0.000$ & $\pm0.000$  & $\pm0.000$\\
$(-0.100, -0.070)$ & $0.003$ & $\pm$0.003 & $\pm0.001$ & $\pm0.000$  & $\pm0.000$\\
$(-0.070, -0.050)$ & $0.003$ & $\pm$0.002 & $\pm0.001$ & $\pm0.001$  & $\pm0.000$\\
$(-0.050, -0.040)$ & $-0.001$ & $\pm$0.004 & $\pm0.001$ & $\pm0.000$  & $\pm0.001$\\
$(0.040, 0.050)$ & $0.000$ & $\pm$0.004 & $\pm0.002$ & $\pm0.001$  & $\pm0.002$\\
$(0.050, 0.070)$ & $0.005$ & $\pm$0.002 & $\pm0.001$ & $\pm0.000$  & $\pm0.001$\\
$(0.070, 0.100)$ & $0.007$ & $\pm$0.003 & $\pm0.002$ & $\pm0.000$  & $\pm0.002$\\
$(0.100, 0.130)$ & $0.025$ & $\pm$0.005 & $\pm0.001$ & $\pm0.001$  & $\pm0.000$\\
$(0.130, 0.200)$ & $0.045$ & $\pm$0.007 & $\pm0.002$ & $\pm0.001$  & $\pm0.002$\\
\end{tabular}
\end{ruledtabular}
\end{minipage}
\begin{minipage}{0.48\linewidth}
\caption{\label{tab:AN_pt_neg_pAl}
Data table of $A_{N}$ for negatively charged hadron ($h^{-}$) in 
transversely-polarized \pAl collisions as a function of $p_T$.}
\begin{ruledtabular}
\begin{tabular}{cccccc}
$p_{T}$(GeV/$c$) & $A_{N}$ & $\delta A_N^{\rm stat}$ & $\delta A_N^{\rm syst}$ & $\delta A_N^{\rm method}$ & $\delta A_N^{\rm smear}$ \\ \hline 
\multicolumn{6}{c}{Forward ($x_F>0$)} \\
(1.50, 2.50) & $-0.006$ & $\pm$0.009 & $\pm0.001$ & $\pm0.001$  & $\pm0.001$\\
(2.50, 3.50) & $-0.009$ & $\pm$0.017 & $\pm0.003$ & $\pm0.002$  & $\pm0.002$\\
(3.50, 7.00) & $-0.005$ & $\pm$0.032 & $\pm0.006$ & $\pm0.002$  & $\pm0.005$\\
\multicolumn{6}{c}{Backward ($x_F<0$)} \\
(1.50, 2.50) & $-0.005$ & $\pm$0.009 & $\pm0.006$ & $\pm0.001$  & $\pm0.006$\\
(2.50, 3.50) & $0.003$ & $\pm$0.019 & $\pm0.004$ & $\pm0.004$  & $\pm0.002$\\
(3.50, 7.00) & $0.013$ & $\pm$0.037 & $\pm0.007$ & $\pm0.002$  & $\pm0.007$\\
\end{tabular}
\end{ruledtabular}
\end{minipage}
\hspace{0.2cm}
\begin{minipage}{0.48\linewidth}
\caption{\label{tab:AN_pt_pos_pAl}
Data table of $A_{N}$ for positively charged hadron ($h^{+}$) in 
transversely-polarized \pAl collisions as a function of $p_T$.}
\begin{ruledtabular}
\begin{tabular}{cccccc}
$p_{T}$(GeV/$c$) & $A_{N}$ & $\delta A_N^{\rm stat}$ & $\delta A_N^{\rm syst}$ & $\delta A_N^{\rm method}$ & $\delta A_N^{\rm smear}$ \\ \hline 
\multicolumn{6}{c}{Forward ($x_F>0$)} \\
(1.50, 2.50) & $0.004$ & $\pm$0.006 & $\pm0.002$ & $\pm0.002$  & $\pm0.000$\\
(2.50, 3.50) & $0.024$ & $\pm$0.011 & $\pm0.006$ & $\pm0.001$  & $\pm0.006$\\
(3.50, 7.00) & $0.029$ & $\pm$0.022 & $\pm0.002$ & $\pm0.001$  & $\pm0.002$\\
\multicolumn{6}{c}{Backward ($x_F<0$)} \\
(1.50, 2.50) & $-0.014$ & $\pm$0.007 & $\pm0.005$ & $\pm0.001$  & $\pm0.005$\\
(2.50, 3.50) & $-0.010$ & $\pm$0.014 & $\pm0.002$ & $\pm0.001$  & $\pm0.002$\\
(3.50, 7.00) & $0.052$ & $\pm$0.028 & $\pm0.021$ & $\pm0.004$  & $\pm0.020$\\
\end{tabular}
\end{ruledtabular}
\end{minipage}
\begin{minipage}{0.48\linewidth}
\caption{\label{tab:AN_xf_neg_pAl}
Data table of $A_{N}$ for negatively charged hadron ($h^{-}$) in 
transversely-polarized \pAl collisions as a function of $x_F$.}
\begin{ruledtabular}
\begin{tabular}{cccccc}
$x_{F}$ & $A_{N}$ & $\delta A_N^{\rm stat}$ & $\delta A_N^{\rm syst}$ & $\delta A_N^{\rm method}$ & $\delta A_N^{\rm smear}$ \\ \hline 
$(-0.200, -0.130)$ & $0.029$ & $\pm$0.040 & $\pm0.009$ & $\pm0.000$  & $\pm0.009$\\
$(-0.130, -0.100)$ & $0.044$ & $\pm$0.026 & $\pm0.013$ & $\pm0.006$  & $\pm0.011$\\
$(-0.100, -0.070)$ & $-0.002$ & $\pm$0.013 & $\pm0.004$ & $\pm0.001$  & $\pm0.004$\\
$(-0.070, -0.050)$ & $-0.005$ & $\pm$0.009 & $\pm0.003$ & $\pm0.001$  & $\pm0.003$\\
$(-0.050, -0.040)$ & $0.009$ & $\pm$0.014 & $\pm0.004$ & $\pm0.000$  & $\pm0.004$\\
$(0.040, 0.050)$ & $0.000$ & $\pm$0.021 & $\pm0.004$ & $\pm0.003$  & $\pm0.003$\\
$(0.050, 0.070)$ & $-0.010$ & $\pm$0.009 & $\pm0.005$ & $\pm0.002$  & $\pm0.004$\\
$(0.070, 0.100)$ & $-0.008$ & $\pm$0.009 & $\pm0.003$ & $\pm0.000$  & $\pm0.003$\\
$(0.100, 0.130)$ & $-0.017$ & $\pm$0.015 & $\pm0.007$ & $\pm0.002$  & $\pm0.007$\\
$(0.130, 0.200)$ & $0.022$ & $\pm$0.020 & $\pm0.003$ & $\pm0.000$  & $\pm0.003$\\
\end{tabular}
\end{ruledtabular}
\end{minipage}
\hspace{0.2cm}
\begin{minipage}{0.48\linewidth}
\caption{\label{tab:AN_xf_pos_pAl}
Data table of $A_{N}$ for positively charged hadron ($h^{+}$) in 
transversely-polarized \pAl collisions as a function of $x_F$.}
\begin{ruledtabular}
\begin{tabular}{cccccc}
$x_{F}$ & $A_{N}$ & $\delta A_N^{\rm stat}$ & $\delta A_N^{\rm syst}$ & $\delta A_N^{\rm method}$ & $\delta A_N^{\rm smear}$ \\ \hline 
$(-0.200, -0.130)$ & $0.074$ & $\pm$0.030 & $\pm0.018$ & $\pm0.005$  & $\pm0.017$\\
$(-0.130, -0.100)$ & $-0.033$ & $\pm$0.020 & $\pm0.014$ & $\pm0.004$  & $\pm0.014$\\
$(-0.100, -0.070)$ & $-0.001$ & $\pm$0.009 & $\pm0.001$ & $\pm0.001$  & $\pm0.001$\\
$(-0.070, -0.050)$ & $-0.014$ & $\pm$0.007 & $\pm0.005$ & $\pm0.001$  & $\pm0.005$\\
$(-0.050, -0.040)$ & $0.002$ & $\pm$0.011 & $\pm0.005$ & $\pm0.002$  & $\pm0.005$\\
$(0.040, 0.050)$ & $0.030$ & $\pm$0.014 & $\pm0.014$ & $\pm0.001$  & $\pm0.014$\\
$(0.050, 0.070)$ & $-0.004$ & $\pm$0.006 & $\pm0.005$ & $\pm0.001$  & $\pm0.004$\\
$(0.070, 0.100)$ & $0.006$ & $\pm$0.006 & $\pm0.001$ & $\pm0.001$  & $\pm0.000$\\
$(0.100, 0.130)$ & $0.010$ & $\pm$0.010 & $\pm0.001$ & $\pm0.001$  & $\pm0.001$\\
$(0.130, 0.200)$ & $0.021$ & $\pm$0.014 & $\pm0.001$ & $\pm0.001$  & $\pm0.000$\\
\end{tabular}
\end{ruledtabular}
\end{minipage}
\end{table*}
\endgroup

\begin{table*}[tbh]
\begin{minipage}{0.48\linewidth}
\caption{\label{tab:AN_pt_neg_pAu}
Data table of $A_{N}$ for negatively charged hadron ($h^{-}$) in 
transversely-polarized \pAu collisions as a function of $p_T$.}
\begin{ruledtabular}
\begin{tabular}{cccccc}
$p_{T}$(GeV/$c$) & $A_{N}$ & $\delta A_N^{\rm stat}$ & $\delta A_N^{\rm syst}$ & $\delta A_N^{\rm method}$ & $\delta A_N^{\rm smear}$ \\ \hline 
\multicolumn{6}{c}{Forward ($x_F>0$)} \\
(1.50, 2.50) & $-0.014$ & $\pm$0.007 & $\pm0.008$ & $\pm0.000$  & $\pm0.008$\\
(2.50, 3.50) & $0.000$ & $\pm$0.013 & $\pm0.003$ & $\pm0.000$  & $\pm0.003$\\
(3.50, 7.00) & $0.012$ & $\pm$0.023 & $\pm0.009$ & $\pm0.005$  & $\pm0.007$\\
\multicolumn{6}{c}{Backward ($x_F<0$)} \\
(1.50, 2.50) & $0.002$ & $\pm$0.006 & $\pm0.003$ & $\pm0.001$  & $\pm0.002$\\
(2.50, 3.50) & $-0.006$ & $\pm$0.013 & $\pm0.003$ & $\pm0.001$  & $\pm0.003$\\
(3.50, 7.00) & $0.003$ & $\pm$0.024 & $\pm0.001$ & $\pm0.000$  & $\pm0.001$\\
\end{tabular}
\end{ruledtabular}
\end{minipage}
\hspace{0.2cm}
\begin{minipage}{0.48\linewidth}
\caption{\label{tab:AN_pt_pos_pAu}
Data table of $A_{N}$ for positively charged hadron ($h^{+}$) in 
transversely-polarized \pAu collisions as a function of $p_T$.}
\begin{ruledtabular}
\begin{tabular}{cccccc}
$p_{T}$(GeV/$c$) & $A_{N}$ & $\delta A_N^{\rm stat}$ & $\delta A_N^{\rm syst}$ & $\delta A_N^{\rm method}$ & $\delta A_N^{\rm smear}$ \\ \hline 
\multicolumn{6}{c}{Forward ($x_F>0$)} \\
(1.50, 2.50) & $0.010$ & $\pm$0.005 & $\pm0.006$ & $\pm0.000$  & $\pm0.006$\\
(2.50, 3.50) & $-0.005$ & $\pm$0.008 & $\pm0.005$ & $\pm0.001$  & $\pm0.005$\\
(3.50, 7.00) & $0.020$ & $\pm$0.016 & $\pm0.008$ & $\pm0.001$  & $\pm0.008$\\
\multicolumn{6}{c}{Backward ($x_F<0$)} \\
(1.50, 2.50) & $0.005$ & $\pm$0.005 & $\pm0.002$ & $\pm0.000$  & $\pm0.002$\\
(2.50, 3.50) & $-0.001$ & $\pm$0.009 & $\pm0.001$ & $\pm0.001$  & $\pm0.001$\\
(3.50, 7.00) & $-0.016$ & $\pm$0.019 & $\pm0.006$ & $\pm0.001$  & $\pm0.006$\\
\end{tabular}
\end{ruledtabular}
\end{minipage}
\begin{minipage}{0.48\linewidth}
\caption{\label{tab:AN_xf_neg_pAu}
Data table of $A_{N}$ for negatively charged hadron ($h^{-}$) in 
transversely-polarized \pAu collisions as a function of $x_F$.}
\begin{ruledtabular}
\begin{tabular}{cccccc}
$x_{F}$ & $A_{N}$ & $\delta A_N^{\rm stat}$ & $\delta A_N^{\rm syst}$ & $\delta A_N^{\rm method}$ & $\delta A_N^{\rm smear}$ \\ \hline 
$(-0.200, -0.130)$ & $-0.013$ & $\pm$0.027 & $\pm0.007$ & $\pm0.000$  & $\pm0.007$\\
$(-0.130, -0.100)$ & $0.012$ & $\pm$0.018 & $\pm0.003$ & $\pm0.000$  & $\pm0.003$\\
$(-0.100, -0.070)$ & $0.001$ & $\pm$0.009 & $\pm0.001$ & $\pm0.000$  & $\pm0.001$\\
$(-0.070, -0.050)$ & $-0.007$ & $\pm$0.006 & $\pm0.004$ & $\pm0.001$  & $\pm0.003$\\
$(-0.050, -0.040)$ & $-0.002$ & $\pm$0.009 & $\pm0.001$ & $\pm0.000$  & $\pm0.001$\\
$(0.040, 0.050)$ & $-0.018$ & $\pm$0.018 & $\pm0.009$ & $\pm0.001$  & $\pm0.009$\\
$(0.050, 0.070)$ & $-0.007$ & $\pm$0.008 & $\pm0.002$ & $\pm0.001$  & $\pm0.002$\\
$(0.070, 0.100)$ & $-0.002$ & $\pm$0.007 & $\pm0.000$ & $\pm0.000$  & $\pm0.000$\\
$(0.100, 0.130)$ & $-0.001$ & $\pm$0.011 & $\pm0.001$ & $\pm0.000$  & $\pm0.000$\\
$(0.130, 0.200)$ & $-0.005$ & $\pm$0.015 & $\pm0.002$ & $\pm0.002$  & $\pm0.000$\\
\end{tabular}
\end{ruledtabular}
\end{minipage}
\hspace{0.2cm}
\begin{minipage}{0.48\linewidth}
\caption{\label{tab:AN_xf_pos_pAu}
Data table of $A_{N}$ for positively charged hadron ($h^{+}$) in 
transversely-polarized \pAu collisions as a function of $x_F$.}
\begin{ruledtabular}
\begin{tabular}{cccccc}
$x_{F}$ & $A_{N}$ & $\delta A_N^{\rm stat}$ & $\delta A_N^{\rm syst}$ & $\delta A_N^{\rm method}$ & $\delta A_N^{\rm smear}$ \\ \hline 
$(-0.200, -0.130)$ & $-0.010$ & $\pm$0.020 & $\pm0.003$ & $\pm0.000$  & $\pm0.003$\\
$(-0.130, -0.100)$ & $0.016$ & $\pm$0.013 & $\pm0.004$ & $\pm0.000$  & $\pm0.004$\\
$(-0.100, -0.070)$ & $0.008$ & $\pm$0.006 & $\pm0.001$ & $\pm0.000$  & $\pm0.001$\\
$(-0.070, -0.050)$ & $-0.004$ & $\pm$0.005 & $\pm0.002$ & $\pm0.000$  & $\pm0.002$\\
$(-0.050, -0.040)$ & $0.000$ & $\pm$0.007 & $\pm0.001$ & $\pm0.000$  & $\pm0.001$\\
$(0.040, 0.050)$ & $0.009$ & $\pm$0.012 & $\pm0.005$ & $\pm0.000$  & $\pm0.004$\\
$(0.050, 0.070)$ & $-0.002$ & $\pm$0.005 & $\pm0.002$ & $\pm0.000$  & $\pm0.002$\\
$(0.070, 0.100)$ & $0.007$ & $\pm$0.004 & $\pm0.002$ & $\pm0.000$  & $\pm0.002$\\
$(0.100, 0.130)$ & $-0.008$ & $\pm$0.008 & $\pm0.004$ & $\pm0.001$  & $\pm0.004$\\
$(0.130, 0.200)$ & $0.017$ & $\pm$0.010 & $\pm0.004$ & $\pm0.001$  & $\pm0.004$\\
\end{tabular}
\end{ruledtabular}
\end{minipage}
\end{table*}

\clearpage



\begin{thebibliography}{60}%
\makeatletter
\providecommand \@ifxundefined [1]{%
 \@ifx{#1\undefined}
}%
\providecommand \@ifnum [1]{%
 \ifnum #1\expandafter \@firstoftwo
 \else \expandafter \@secondoftwo
 \fi
}%
\providecommand \@ifx [1]{%
 \ifx #1\expandafter \@firstoftwo
 \else \expandafter \@secondoftwo
 \fi
}%
\providecommand \natexlab [1]{#1}%
\providecommand \enquote  [1]{``#1''}%
\providecommand \bibnamefont  [1]{#1}%
\providecommand \bibfnamefont [1]{#1}%
\providecommand \citenamefont [1]{#1}%
\providecommand \href@noop [0]{\@secondoftwo}%
\providecommand \href [0]{\begingroup \@sanitize@url \@href}%
\providecommand \@href[1]{\@@startlink{#1}\@@href}%
\providecommand \@@href[1]{\endgroup#1\@@endlink}%
\providecommand \@sanitize@url [0]{\catcode `\\12\catcode `\$12\catcode
  `\&12\catcode `\#12\catcode `\^12\catcode `\_12\catcode `\%12\relax}%
\providecommand \@@startlink[1]{}%
\providecommand \@@endlink[0]{}%
\providecommand \url  [0]{\begingroup\@sanitize@url \@url }%
\providecommand \@url [1]{\endgroup\@href {#1}{\urlprefix }}%
\providecommand \urlprefix  [0]{URL }%
\providecommand \Eprint [0]{\href }%
\providecommand \doibase [0]{https://doi.org/}%
\providecommand \selectlanguage [0]{\@gobble}%
\providecommand \bibinfo  [0]{\@secondoftwo}%
\providecommand \bibfield  [0]{\@secondoftwo}%
\providecommand \translation [1]{[#1]}%
\providecommand \BibitemOpen [0]{}%
\providecommand \bibitemStop [0]{}%
\providecommand \bibitemNoStop [0]{.\EOS\space}%
\providecommand \EOS [0]{\spacefactor3000\relax}%
\providecommand \BibitemShut  [1]{\csname bibitem#1\endcsname}%
\let\auto@bib@innerbib\@empty
\bibitem [{\citenamefont {Klem}\ \emph {et~al.}(1976)\citenamefont {Klem},
  \citenamefont {Bowers}, \citenamefont {Courant}, \citenamefont {Kagan},
  \citenamefont {Marshak}, \citenamefont {Peterson}, \citenamefont {Ruddick},
  \citenamefont {Dragoset},\ and\ \citenamefont {Roberts}}]{Klem:1976ui}%
  \BibitemOpen
  \bibfield  {author} {\bibinfo {author} {\bibfnamefont {R.~D.}\ \bibnamefont
  {Klem}}, \bibinfo {author} {\bibfnamefont {J.~E.}\ \bibnamefont {Bowers}},
  \bibinfo {author} {\bibfnamefont {H.~W.}\ \bibnamefont {Courant}}, \bibinfo
  {author} {\bibfnamefont {H.}~\bibnamefont {Kagan}}, \bibinfo {author}
  {\bibfnamefont {M.~L.}\ \bibnamefont {Marshak}}, \bibinfo {author}
  {\bibfnamefont {E.~A.}\ \bibnamefont {Peterson}}, \bibinfo {author}
  {\bibfnamefont {K.}~\bibnamefont {Ruddick}}, \bibinfo {author} {\bibfnamefont
  {W.~H.}\ \bibnamefont {Dragoset}},\ and\ \bibinfo {author} {\bibfnamefont
  {J.~B.}\ \bibnamefont {Roberts}},\ }\bibfield  {title} {\bibinfo {title}
  {{Measurement of Asymmetries of Inclusive Pion Production in Proton Proton
  Interactions at 6 GeV/$c$ and 11.8 GeV/$c$}},\ }\href
  {https://doi.org/10.1103/PhysRevLett.36.929} {\bibfield  {journal} {\bibinfo
  {journal} {Phys. Rev. Lett.}\ }\textbf {\bibinfo {volume} {36}},\ \bibinfo
  {pages} {929} (\bibinfo {year} {1976})}\BibitemShut {NoStop}%
\bibitem [{\citenamefont {Antille}\ \emph {et~al.}(1980)\citenamefont
  {Antille}, \citenamefont {Dick}, \citenamefont {Madansky}, \citenamefont
  {Perret-Gallix}, \citenamefont {Werlen}, \citenamefont {Gonidec},
  \citenamefont {Kuroda},\ and\ \citenamefont {Kyberd}}]{Antille:1980th}%
  \BibitemOpen
  \bibfield  {author} {\bibinfo {author} {\bibfnamefont {J.}~\bibnamefont
  {Antille}}, \bibinfo {author} {\bibfnamefont {L.}~\bibnamefont {Dick}},
  \bibinfo {author} {\bibfnamefont {L.}~\bibnamefont {Madansky}}, \bibinfo
  {author} {\bibfnamefont {D.}~\bibnamefont {Perret-Gallix}}, \bibinfo {author}
  {\bibfnamefont {M.}~\bibnamefont {Werlen}}, \bibinfo {author} {\bibfnamefont
  {A.}~\bibnamefont {Gonidec}}, \bibinfo {author} {\bibfnamefont
  {K.}~\bibnamefont {Kuroda}},\ and\ \bibinfo {author} {\bibfnamefont
  {P.}~\bibnamefont {Kyberd}},\ }\bibfield  {title} {\bibinfo {title} {{Spin
  Dependence of the Inclusive Reaction $p p$ (Polarized) $\to \pi^0 X$ at 24
  {GeV}/$c$ for High $p_T$ $\pi^0$ Produced in the Central Region}},\ }\href
  {https://doi.org/10.1016/0370-2693(80)90933-8} {\bibfield  {journal}
  {\bibinfo  {journal} {Phys. Lett. B}\ }\textbf {\bibinfo {volume} {94}},\
  \bibinfo {pages} {523} (\bibinfo {year} {1980})}\BibitemShut {NoStop}%
\bibitem [{\citenamefont {Adams}\ \emph
  {et~al.}(1991{\natexlab{a}})\citenamefont {Adams} \emph
  {et~al.}}]{FNAL-E704:1991ovg}%
  \BibitemOpen
  \bibfield  {author} {\bibinfo {author} {\bibfnamefont {D.~L.}\ \bibnamefont
  {Adams}} \emph {et~al.} (\bibinfo {collaboration} {FNAL-E704
  Collaboration}),\ }\bibfield  {title} {\bibinfo {title} {{Analyzing power in
  inclusive $\pi^{+}$ and $\pi^{-}$ production at high $x_{\mathrm{F}}$ with a
  200 GeV polarized proton beam}},\ }\href
  {https://doi.org/10.1016/0370-2693(91)90378-4} {\bibfield  {journal}
  {\bibinfo  {journal} {Phys. Lett. B}\ }\textbf {\bibinfo {volume} {264}},\
  \bibinfo {pages} {462} (\bibinfo {year} {1991}{\natexlab{a}})}\BibitemShut
  {NoStop}%
\bibitem [{\citenamefont {Adams}\ \emph
  {et~al.}(1991{\natexlab{b}})\citenamefont {Adams} \emph
  {et~al.}}]{E581:1991eys}%
  \BibitemOpen
  \bibfield  {author} {\bibinfo {author} {\bibfnamefont {D.~L.}\ \bibnamefont
  {Adams}} \emph {et~al.} (\bibinfo {collaboration} {E581, E704
  Collaboration}),\ }\bibfield  {title} {\bibinfo {title} {{Comparison of spin
  asymmetries and cross-sections in $\pi^{0}$ production by 200 GeV polarized
  antiprotons and protons}},\ }\href
  {https://doi.org/10.1016/0370-2693(91)91351-U} {\bibfield  {journal}
  {\bibinfo  {journal} {Phys. Lett. B}\ }\textbf {\bibinfo {volume} {261}},\
  \bibinfo {pages} {201} (\bibinfo {year} {1991}{\natexlab{b}})}\BibitemShut
  {NoStop}%
\bibitem [{\citenamefont {Allgower}\ \emph {et~al.}(2002)\citenamefont
  {Allgower}, \citenamefont {Krueger}, \citenamefont {Kasprzyk}, \citenamefont
  {Spinka}, \citenamefont {Underwood}, \citenamefont {Yokosawa} \emph
  {et~al.}}]{Allgower:2002qi}%
  \BibitemOpen
  \bibfield  {author} {\bibinfo {author} {\bibfnamefont {C.~E.}\ \bibnamefont
  {Allgower}}, \bibinfo {author} {\bibfnamefont {K.~W.}\ \bibnamefont
  {Krueger}}, \bibinfo {author} {\bibfnamefont {T.~E.}\ \bibnamefont
  {Kasprzyk}}, \bibinfo {author} {\bibfnamefont {H.~M.}\ \bibnamefont
  {Spinka}}, \bibinfo {author} {\bibfnamefont {D.~G.}\ \bibnamefont
  {Underwood}}, \bibinfo {author} {\bibfnamefont {A.}~\bibnamefont {Yokosawa}},
  \emph {et~al.},\ }\bibfield  {title} {\bibinfo {title} {{Measurement of
  analyzing powers of $\pi^{+}$ and $\pi^{-}$ produced on a hydrogen and a
  carbon target with a 22-GeV/$c$ incident polarized proton beam}},\ }\href
  {https://doi.org/10.1103/PhysRevD.65.092008} {\bibfield  {journal} {\bibinfo
  {journal} {Phys. Rev. D}\ }\textbf {\bibinfo {volume} {65}},\ \bibinfo
  {pages} {092008} (\bibinfo {year} {2002})}\BibitemShut {NoStop}%
\bibitem [{\citenamefont {Adams}\ \emph {et~al.}(2004)\citenamefont {Adams}
  \emph {et~al.}}]{STAR:2003lxu}%
  \BibitemOpen
  \bibfield  {author} {\bibinfo {author} {\bibfnamefont {J.}~\bibnamefont
  {Adams}} \emph {et~al.} (\bibinfo {collaboration} {STAR Collaboration}),\
  }\bibfield  {title} {\bibinfo {title} {{Cross-sections and transverse
  single-spin asymmetries in forward neutral pion production from proton
  collisions at $\sqrt{s}$ = 200 GeV}},\ }\href
  {https://doi.org/10.1103/PhysRevLett.92.171801} {\bibfield  {journal}
  {\bibinfo  {journal} {Phys. Rev. Lett.}\ }\textbf {\bibinfo {volume} {92}},\
  \bibinfo {pages} {171801} (\bibinfo {year} {2004})}\BibitemShut {NoStop}%
\bibitem [{\citenamefont {Lee}\ and\ \citenamefont
  {Videbaek}(2007)}]{Lee:2007zzh}%
  \BibitemOpen
  \bibfield  {author} {\bibinfo {author} {\bibfnamefont {J.~H.}\ \bibnamefont
  {Lee}}\ and\ \bibinfo {author} {\bibfnamefont {F.}~\bibnamefont {Videbaek}}
  (\bibinfo {collaboration} {BRAHMS Collaboration}),\ }\bibfield  {title}
  {\bibinfo {title} {{Single-spin asymmetries of identified hadrons in
  polarized p+p at $\sqrt{s}$ = 62.4 and 200 GeV}},\ }\href
  {https://doi.org/10.1063/1.2750837} {\bibfield  {journal} {\bibinfo
  {journal} {AIP Conf. Proc.}\ }\textbf {\bibinfo {volume} {915}},\ \bibinfo
  {pages} {533} (\bibinfo {year} {2007})}\BibitemShut {NoStop}%
\bibitem [{\citenamefont {Abelev}\ \emph {et~al.}(2008)\citenamefont {Abelev}
  \emph {et~al.}}]{STAR:2008ixi}%
  \BibitemOpen
  \bibfield  {author} {\bibinfo {author} {\bibfnamefont {B.~I.}\ \bibnamefont
  {Abelev}} \emph {et~al.} (\bibinfo {collaboration} {STAR Collaboration}),\
  }\bibfield  {title} {\bibinfo {title} {{Forward Neutral Pion Transverse
  Single-Spin Asymmetries in p+p Collisions at $\sqrt{s}$ = 200 GeV}},\ }\href
  {https://doi.org/10.1103/PhysRevLett.101.222001} {\bibfield  {journal}
  {\bibinfo  {journal} {Phys. Rev. Lett.}\ }\textbf {\bibinfo {volume} {101}},\
  \bibinfo {pages} {222001} (\bibinfo {year} {2008})}\BibitemShut {NoStop}%
\bibitem [{\citenamefont {Arsene}\ \emph {et~al.}(2008)\citenamefont {Arsene}
  \emph {et~al.}}]{BRAHMS:2008doi}%
  \BibitemOpen
  \bibfield  {author} {\bibinfo {author} {\bibfnamefont {I.}~\bibnamefont
  {Arsene}} \emph {et~al.} (\bibinfo {collaboration} {BRAHMS Collaboration}),\
  }\bibfield  {title} {\bibinfo {title} {{Single Transverse Spin Asymmetries of
  Identified Charged Hadrons in Polarized $pp$ Collisions at $\sqrt{s}$ = 62.4
  GeV}},\ }\href {https://doi.org/10.1103/PhysRevLett.101.042001} {\bibfield
  {journal} {\bibinfo  {journal} {Phys. Rev. Lett.}\ }\textbf {\bibinfo
  {volume} {101}},\ \bibinfo {pages} {042001} (\bibinfo {year}
  {2008})}\BibitemShut {NoStop}%
\bibitem [{\citenamefont {Adamczyk}\ \emph {et~al.}(2012)\citenamefont
  {Adamczyk} \emph {et~al.}}]{STAR:2012ljf}%
  \BibitemOpen
  \bibfield  {author} {\bibinfo {author} {\bibfnamefont {L.}~\bibnamefont
  {Adamczyk}} \emph {et~al.} (\bibinfo {collaboration} {STAR Collaboration}),\
  }\bibfield  {title} {\bibinfo {title} {{Transverse Single-Spin Asymmetry and
  Cross-Section for $\pi^0$ and $\eta$ Mesons at Large Feynman-$x$ in Polarized
  $p+p$ Collisions at $\sqrt{s}=200$ GeV}},\ }\href
  {https://doi.org/10.1103/PhysRevD.86.051101} {\bibfield  {journal} {\bibinfo
  {journal} {Phys. Rev. D}\ }\textbf {\bibinfo {volume} {86}},\ \bibinfo
  {pages} {051101} (\bibinfo {year} {2012})}\BibitemShut {NoStop}%
\bibitem [{\citenamefont {Adare}\ \emph
  {et~al.}(2014{\natexlab{a}})\citenamefont {Adare} \emph
  {et~al.}}]{PHENIX:2013wle}%
  \BibitemOpen
  \bibfield  {author} {\bibinfo {author} {\bibfnamefont {A.}~\bibnamefont
  {Adare}} \emph {et~al.} (\bibinfo {collaboration} {PHENIX Collaboration}),\
  }\bibfield  {title} {\bibinfo {title} {{Measurement of transverse-single-spin
  asymmetries for midrapidity and forward-rapidity production of hadrons in
  polarized p+p collisions at $\sqrt{s}=$200 and 62.4 GeV}},\ }\href
  {https://doi.org/10.1103/PhysRevD.90.012006} {\bibfield  {journal} {\bibinfo
  {journal} {Phys. Rev. D}\ }\textbf {\bibinfo {volume} {90}},\ \bibinfo
  {pages} {012006} (\bibinfo {year} {2014}{\natexlab{a}})}\BibitemShut
  {NoStop}%
\bibitem [{\citenamefont {Adare}\ \emph
  {et~al.}(2014{\natexlab{b}})\citenamefont {Adare} \emph
  {et~al.}}]{PHENIX:2014qwb}%
  \BibitemOpen
  \bibfield  {author} {\bibinfo {author} {\bibfnamefont {A.}~\bibnamefont
  {Adare}} \emph {et~al.} (\bibinfo {collaboration} {PHENIX Collaboration}),\
  }\bibfield  {title} {\bibinfo {title} {{Cross section and transverse
  single-spin asymmetry of $\eta$ mesons in $p^{\uparrow}+p$ collisions at
  $\sqrt{s}=200$ GeV at forward rapidity}},\ }\href
  {https://doi.org/10.1103/PhysRevD.90.072008} {\bibfield  {journal} {\bibinfo
  {journal} {Phys. Rev. D}\ }\textbf {\bibinfo {volume} {90}},\ \bibinfo
  {pages} {072008} (\bibinfo {year} {2014}{\natexlab{b}})}\BibitemShut
  {NoStop}%
\bibitem [{\citenamefont {Adam}\ \emph
  {et~al.}(2021{\natexlab{a}})\citenamefont {Adam} \emph
  {et~al.}}]{STAR:2020nnl}%
  \BibitemOpen
  \bibfield  {author} {\bibinfo {author} {\bibfnamefont {J.}~\bibnamefont
  {Adam}} \emph {et~al.} (\bibinfo {collaboration} {STAR Collaboration}),\
  }\bibfield  {title} {\bibinfo {title} {{Measurement of transverse single-spin
  asymmetries of $\pi^0$ and electromagnetic jets at forward rapidity in 200
  and 500 GeV transversely polarized proton-proton collisions}},\ }\href
  {https://doi.org/10.1103/PhysRevD.103.092009} {\bibfield  {journal} {\bibinfo
   {journal} {Phys. Rev. D}\ }\textbf {\bibinfo {volume} {103}},\ \bibinfo
  {pages} {092009} (\bibinfo {year} {2021}{\natexlab{a}})}\BibitemShut
  {NoStop}%
\bibitem [{\citenamefont {Sivers}(1990)}]{Sivers:1989cc}%
  \BibitemOpen
  \bibfield  {author} {\bibinfo {author} {\bibfnamefont {D.~W.}\ \bibnamefont
  {Sivers}},\ }\bibfield  {title} {\bibinfo {title} {{Single-Spin Production
  Asymmetries from the Hard Scattering of Point-Like Constituents}},\ }\href
  {https://doi.org/10.1103/PhysRevD.41.83} {\bibfield  {journal} {\bibinfo
  {journal} {Phys. Rev. D}\ }\textbf {\bibinfo {volume} {41}},\ \bibinfo
  {pages} {83} (\bibinfo {year} {1990})}\BibitemShut {NoStop}%
\bibitem [{\citenamefont {Sivers}(1991)}]{Sivers:1990fh}%
  \BibitemOpen
  \bibfield  {author} {\bibinfo {author} {\bibfnamefont {D.~W.}\ \bibnamefont
  {Sivers}},\ }\bibfield  {title} {\bibinfo {title} {{Hard scattering scaling
  laws for single-spin production asymmetries}},\ }\href
  {https://doi.org/10.1103/PhysRevD.43.261} {\bibfield  {journal} {\bibinfo
  {journal} {Phys. Rev. D}\ }\textbf {\bibinfo {volume} {43}},\ \bibinfo
  {pages} {261} (\bibinfo {year} {1991})}\BibitemShut {NoStop}%
\bibitem [{\citenamefont {Collins}(1993)}]{Collins:1992kk}%
  \BibitemOpen
  \bibfield  {author} {\bibinfo {author} {\bibfnamefont {J.~C.}\ \bibnamefont
  {Collins}},\ }\bibfield  {title} {\bibinfo {title} {{Fragmentation of
  transversely polarized quarks probed in transverse momentum distributions}},\
  }\href {https://doi.org/10.1016/0550-3213(93)90262-N} {\bibfield  {journal}
  {\bibinfo  {journal} {Nucl. Phys. B}\ }\textbf {\bibinfo {volume} {396}},\
  \bibinfo {pages} {161} (\bibinfo {year} {1993})}\BibitemShut {NoStop}%
\bibitem [{\citenamefont {Qiu}\ and\ \citenamefont
  {Sterman}(1998)}]{Qiu:1998ia}%
  \BibitemOpen
  \bibfield  {author} {\bibinfo {author} {\bibfnamefont {J.}~\bibnamefont
  {Qiu}}\ and\ \bibinfo {author} {\bibfnamefont {G.~F.}\ \bibnamefont
  {Sterman}},\ }\bibfield  {title} {\bibinfo {title} {{Single transverse spin
  asymmetries in hadronic pion production}},\ }\href
  {https://doi.org/10.1103/PhysRevD.59.014004} {\bibfield  {journal} {\bibinfo
  {journal} {Phys. Rev. D}\ }\textbf {\bibinfo {volume} {59}},\ \bibinfo
  {pages} {014004} (\bibinfo {year} {1998})}\BibitemShut {NoStop}%
\bibitem [{\citenamefont {Kouvaris}\ \emph {et~al.}(2006)\citenamefont
  {Kouvaris}, \citenamefont {Qiu}, \citenamefont {Vogelsang},\ and\
  \citenamefont {Yuan}}]{Kouvaris:2006zy}%
  \BibitemOpen
  \bibfield  {author} {\bibinfo {author} {\bibfnamefont {C.}~\bibnamefont
  {Kouvaris}}, \bibinfo {author} {\bibfnamefont {J.~W.}\ \bibnamefont {Qiu}},
  \bibinfo {author} {\bibfnamefont {W.}~\bibnamefont {Vogelsang}},\ and\
  \bibinfo {author} {\bibfnamefont {F.}~\bibnamefont {Yuan}},\ }\bibfield
  {title} {\bibinfo {title} {{Single transverse-spin asymmetry in high
  transverse momentum pion production in pp collisions}},\ }\href
  {https://doi.org/10.1103/PhysRevD.74.114013} {\bibfield  {journal} {\bibinfo
  {journal} {Phys. Rev. D}\ }\textbf {\bibinfo {volume} {74}},\ \bibinfo
  {pages} {114013} (\bibinfo {year} {2006})}\BibitemShut {NoStop}%
\bibitem [{\citenamefont {Koike}\ and\ \citenamefont
  {Tanaka}(2007)}]{Koike:2007rq}%
  \BibitemOpen
  \bibfield  {author} {\bibinfo {author} {\bibfnamefont {Y.}~\bibnamefont
  {Koike}}\ and\ \bibinfo {author} {\bibfnamefont {K.}~\bibnamefont {Tanaka}},\
  }\bibfield  {title} {\bibinfo {title} {{Universal structure of twist-3
  soft-gluon-pole cross-sections for single transverse-spin asymmetry}},\
  }\href {https://doi.org/10.1103/PhysRevD.76.011502} {\bibfield  {journal}
  {\bibinfo  {journal} {Phys. Rev. D}\ }\textbf {\bibinfo {volume} {76}},\
  \bibinfo {pages} {011502(R)} (\bibinfo {year} {2007})}\BibitemShut {NoStop}%
\bibitem [{\citenamefont {Koike}\ and\ \citenamefont
  {Tomita}(2009)}]{Koike:2009ge}%
  \BibitemOpen
  \bibfield  {author} {\bibinfo {author} {\bibfnamefont {Y.}~\bibnamefont
  {Koike}}\ and\ \bibinfo {author} {\bibfnamefont {T.}~\bibnamefont {Tomita}},\
  }\bibfield  {title} {\bibinfo {title} {{Soft-fermion-pole contribution to
  single-spin asymmetry for pion production in pp collisions}},\ }\href
  {https://doi.org/10.1016/j.physletb.2009.04.017} {\bibfield  {journal}
  {\bibinfo  {journal} {Phys. Lett. B}\ }\textbf {\bibinfo {volume} {675}},\
  \bibinfo {pages} {181} (\bibinfo {year} {2009})}\BibitemShut {NoStop}%
\bibitem [{\citenamefont {Kanazawa}\ and\ \citenamefont
  {Koike}(2010)}]{Kanazawa:2010au}%
  \BibitemOpen
  \bibfield  {author} {\bibinfo {author} {\bibfnamefont {K.}~\bibnamefont
  {Kanazawa}}\ and\ \bibinfo {author} {\bibfnamefont {Y.}~\bibnamefont
  {Koike}},\ }\bibfield  {title} {\bibinfo {title} {{New Analysis of the Single
  Transverse-Spin Asymmetry for Hadron Production at RHIC}},\ }\href
  {https://doi.org/10.1103/PhysRevD.82.034009} {\bibfield  {journal} {\bibinfo
  {journal} {Phys. Rev. D}\ }\textbf {\bibinfo {volume} {82}},\ \bibinfo
  {pages} {034009} (\bibinfo {year} {2010})}\BibitemShut {NoStop}%
\bibitem [{\citenamefont {Kang}\ \emph {et~al.}(2011)\citenamefont {Kang},
  \citenamefont {Qiu}, \citenamefont {Vogelsang},\ and\ \citenamefont
  {Yuan}}]{Kang:2011hk}%
  \BibitemOpen
  \bibfield  {author} {\bibinfo {author} {\bibfnamefont {Z.~B.}\ \bibnamefont
  {Kang}}, \bibinfo {author} {\bibfnamefont {J.~W.}\ \bibnamefont {Qiu}},
  \bibinfo {author} {\bibfnamefont {W.}~\bibnamefont {Vogelsang}},\ and\
  \bibinfo {author} {\bibfnamefont {F.}~\bibnamefont {Yuan}},\ }\bibfield
  {title} {\bibinfo {title} {{An Observation Concerning the Process Dependence
  of the Sivers Functions}},\ }\href
  {https://doi.org/10.1103/PhysRevD.83.094001} {\bibfield  {journal} {\bibinfo
  {journal} {Phys. Rev. D}\ }\textbf {\bibinfo {volume} {83}},\ \bibinfo
  {pages} {094001} (\bibinfo {year} {2011})}\BibitemShut {NoStop}%
\bibitem [{\citenamefont {Kanazawa}\ and\ \citenamefont
  {Koike}(2011)}]{Kanazawa:2011bg}%
  \BibitemOpen
  \bibfield  {author} {\bibinfo {author} {\bibfnamefont {K.}~\bibnamefont
  {Kanazawa}}\ and\ \bibinfo {author} {\bibfnamefont {Y.}~\bibnamefont
  {Koike}},\ }\bibfield  {title} {\bibinfo {title} {{A phenomenological study
  on single transverse-spin asymmetry for inclusive light-hadron productions at
  RHIC}},\ }\href {https://doi.org/10.1103/PhysRevD.83.114024} {\bibfield
  {journal} {\bibinfo  {journal} {Phys. Rev. D}\ }\textbf {\bibinfo {volume}
  {83}},\ \bibinfo {pages} {114024} (\bibinfo {year} {2011})}\BibitemShut
  {NoStop}%
\bibitem [{\citenamefont {Kang}\ and\ \citenamefont
  {Prokudin}(2012)}]{Kang:2012xf}%
  \BibitemOpen
  \bibfield  {author} {\bibinfo {author} {\bibfnamefont {Z.~B.}\ \bibnamefont
  {Kang}}\ and\ \bibinfo {author} {\bibfnamefont {A.}~\bibnamefont
  {Prokudin}},\ }\bibfield  {title} {\bibinfo {title} {{Global fitting of
  single-spin asymmetry: an attempt}},\ }\href
  {https://doi.org/10.1103/PhysRevD.85.074008} {\bibfield  {journal} {\bibinfo
  {journal} {Phys. Rev. D}\ }\textbf {\bibinfo {volume} {85}},\ \bibinfo
  {pages} {074008} (\bibinfo {year} {2012})}\BibitemShut {NoStop}%
\bibitem [{\citenamefont {Beppu}\ \emph {et~al.}(2014)\citenamefont {Beppu},
  \citenamefont {Kanazawa}, \citenamefont {Koike},\ and\ \citenamefont
  {Yoshida}}]{Beppu:2013uda}%
  \BibitemOpen
  \bibfield  {author} {\bibinfo {author} {\bibfnamefont {H.}~\bibnamefont
  {Beppu}}, \bibinfo {author} {\bibfnamefont {K.}~\bibnamefont {Kanazawa}},
  \bibinfo {author} {\bibfnamefont {Y.}~\bibnamefont {Koike}},\ and\ \bibinfo
  {author} {\bibfnamefont {S.}~\bibnamefont {Yoshida}},\ }\bibfield  {title}
  {\bibinfo {title} {{Three-gluon contribution to the single-spin asymmetry for
  light hadron production in pp collision}},\ }\href
  {https://doi.org/10.1103/PhysRevD.89.034029} {\bibfield  {journal} {\bibinfo
  {journal} {Phys. Rev. D}\ }\textbf {\bibinfo {volume} {89}},\ \bibinfo
  {pages} {034029} (\bibinfo {year} {2014})}\BibitemShut {NoStop}%
\bibitem [{\citenamefont {Metz}\ and\ \citenamefont
  {Pitonyak}(2013)}]{Metz:2012ct}%
  \BibitemOpen
  \bibfield  {author} {\bibinfo {author} {\bibfnamefont {A.}~\bibnamefont
  {Metz}}\ and\ \bibinfo {author} {\bibfnamefont {D.}~\bibnamefont
  {Pitonyak}},\ }\bibfield  {title} {\bibinfo {title} {{Fragmentation
  contribution to the transverse single-spin asymmetry in proton-proton
  collisions}},\ }\href {https://doi.org/10.1016/j.physletb.2013.05.043}
  {\bibfield  {journal} {\bibinfo  {journal} {Phys. Lett. B}\ }\textbf
  {\bibinfo {volume} {723}},\ \bibinfo {pages} {365} (\bibinfo {year}
  {2013})},\ \bibinfo {note} {[Phys. Lett. B {\bf 762}, 549(E)
  (2016)]}\BibitemShut {NoStop}%
\bibitem [{\citenamefont {Kanazawa}\ \emph {et~al.}(2014)\citenamefont
  {Kanazawa}, \citenamefont {Koike}, \citenamefont {Metz},\ and\ \citenamefont
  {Pitonyak}}]{Kanazawa:2014dca}%
  \BibitemOpen
  \bibfield  {author} {\bibinfo {author} {\bibfnamefont {K.}~\bibnamefont
  {Kanazawa}}, \bibinfo {author} {\bibfnamefont {Y.}~\bibnamefont {Koike}},
  \bibinfo {author} {\bibfnamefont {A.}~\bibnamefont {Metz}},\ and\ \bibinfo
  {author} {\bibfnamefont {D.}~\bibnamefont {Pitonyak}},\ }\bibfield  {title}
  {\bibinfo {title} {{Towards an explanation of transverse single-spin
  asymmetries in proton-proton collisions: the role of fragmentation in
  collinear factorization}},\ }\href
  {https://doi.org/10.1103/PhysRevD.89.111501} {\bibfield  {journal} {\bibinfo
  {journal} {Phys. Rev. D}\ }\textbf {\bibinfo {volume} {89}},\ \bibinfo
  {pages} {111501(R)} (\bibinfo {year} {2014})}\BibitemShut {NoStop}%
\bibitem [{\citenamefont {Gamberg}\ \emph {et~al.}(2017)\citenamefont
  {Gamberg}, \citenamefont {Kang}, \citenamefont {Pitonyak},\ and\
  \citenamefont {Prokudin}}]{Gamberg:2017gle}%
  \BibitemOpen
  \bibfield  {author} {\bibinfo {author} {\bibfnamefont {L.}~\bibnamefont
  {Gamberg}}, \bibinfo {author} {\bibfnamefont {Z.}~\bibnamefont {Kang}},
  \bibinfo {author} {\bibfnamefont {D.}~\bibnamefont {Pitonyak}},\ and\
  \bibinfo {author} {\bibfnamefont {A.}~\bibnamefont {Prokudin}},\ }\bibfield
  {title} {\bibinfo {title} {{Phenomenological constraints on $A_N$ in
  $p^\uparrow p\to \pi\, X$ from Lorentz invariance relations}},\ }\href
  {https://doi.org/10.1016/j.physletb.2017.04.061} {\bibfield  {journal}
  {\bibinfo  {journal} {Phys. Lett. B}\ }\textbf {\bibinfo {volume} {770}},\
  \bibinfo {pages} {242} (\bibinfo {year} {2017})}\BibitemShut {NoStop}%
\bibitem [{\citenamefont {Cammarota}\ \emph {et~al.}(2020)\citenamefont
  {Cammarota}, \citenamefont {Gamberg}, \citenamefont {Kang}, \citenamefont
  {Miller}, \citenamefont {Pitonyak}, \citenamefont {Prokudin}, \citenamefont
  {Rogers},\ and\ \citenamefont {Sato}}]{Cammarota:2020qcw}%
  \BibitemOpen
  \bibfield  {author} {\bibinfo {author} {\bibfnamefont {J.}~\bibnamefont
  {Cammarota}}, \bibinfo {author} {\bibfnamefont {L.}~\bibnamefont {Gamberg}},
  \bibinfo {author} {\bibfnamefont {Z.~B.}\ \bibnamefont {Kang}}, \bibinfo
  {author} {\bibfnamefont {J.~A.}\ \bibnamefont {Miller}}, \bibinfo {author}
  {\bibfnamefont {D.}~\bibnamefont {Pitonyak}}, \bibinfo {author}
  {\bibfnamefont {A.}~\bibnamefont {Prokudin}}, \bibinfo {author}
  {\bibfnamefont {T.~C.}\ \bibnamefont {Rogers}},\ and\ \bibinfo {author}
  {\bibfnamefont {N.}~\bibnamefont {Sato}} (\bibinfo {collaboration} {Jefferson
  Lab Angular Momentum Collaboration}),\ }\bibfield  {title} {\bibinfo {title}
  {{Origin of single transverse-spin asymmetries in high-energy collisions}},\
  }\href {https://doi.org/10.1103/PhysRevD.102.054002} {\bibfield  {journal}
  {\bibinfo  {journal} {Phys. Rev. D}\ }\textbf {\bibinfo {volume} {102}},\
  \bibinfo {pages} {054002} (\bibinfo {year} {2020})}\BibitemShut {NoStop}%
\bibitem [{\citenamefont {Gelis}\ \emph {et~al.}(2010)\citenamefont {Gelis},
  \citenamefont {Iancu}, \citenamefont {Jalilian-Marian},\ and\ \citenamefont
  {Venugopalan}}]{Gelis:2010nm}%
  \BibitemOpen
  \bibfield  {author} {\bibinfo {author} {\bibfnamefont {F.}~\bibnamefont
  {Gelis}}, \bibinfo {author} {\bibfnamefont {E.}~\bibnamefont {Iancu}},
  \bibinfo {author} {\bibfnamefont {J.}~\bibnamefont {Jalilian-Marian}},\ and\
  \bibinfo {author} {\bibfnamefont {R.}~\bibnamefont {Venugopalan}},\
  }\bibfield  {title} {\bibinfo {title} {{The Color Glass Condensate}},\ }\href
  {https://doi.org/10.1146/annurev.nucl.010909.083629} {\bibfield  {journal}
  {\bibinfo  {journal} {Ann. Rev. Nucl. Part. Sci.}\ }\textbf {\bibinfo
  {volume} {60}},\ \bibinfo {pages} {463} (\bibinfo {year} {2010})}\BibitemShut
  {NoStop}%
\bibitem [{\citenamefont {Kang}\ and\ \citenamefont
  {Yuan}(2011)}]{Kang:2011ni}%
  \BibitemOpen
  \bibfield  {author} {\bibinfo {author} {\bibfnamefont {Z.~B.}\ \bibnamefont
  {Kang}}\ and\ \bibinfo {author} {\bibfnamefont {F.}~\bibnamefont {Yuan}},\
  }\bibfield  {title} {\bibinfo {title} {{Single-Spin Asymmetry Scaling in the
  Forward Rapidity Region at RHIC}},\ }\href
  {https://doi.org/10.1103/PhysRevD.84.034019} {\bibfield  {journal} {\bibinfo
  {journal} {Phys. Rev. D}\ }\textbf {\bibinfo {volume} {84}},\ \bibinfo
  {pages} {034019} (\bibinfo {year} {2011})}\BibitemShut {NoStop}%
\bibitem [{\citenamefont {Kovchegov}\ and\ \citenamefont
  {Sievert}(2012)}]{Kovchegov:2012ga}%
  \BibitemOpen
  \bibfield  {author} {\bibinfo {author} {\bibfnamefont {Y.~V.}\ \bibnamefont
  {Kovchegov}}\ and\ \bibinfo {author} {\bibfnamefont {M.~D.}\ \bibnamefont
  {Sievert}},\ }\bibfield  {title} {\bibinfo {title} {{A New Mechanism for
  Generating a Single Transverse Spin Asymmetry}},\ }\href
  {https://doi.org/10.1103/PhysRevD.86.034028} {\bibfield  {journal} {\bibinfo
  {journal} {Phys. Rev. D}\ }\textbf {\bibinfo {volume} {86}},\ \bibinfo
  {pages} {034028} (\bibinfo {year} {2012})},\ \bibinfo {note} {[Phys. Rev. D
  {\bf 86}, 079906(E) (2012)]}\BibitemShut {NoStop}%
\bibitem [{\citenamefont {Sch\"afer}\ and\ \citenamefont
  {Zhou}(2014)}]{Schafer:2014zea}%
  \BibitemOpen
  \bibfield  {author} {\bibinfo {author} {\bibfnamefont {A.}~\bibnamefont
  {Sch\"afer}}\ and\ \bibinfo {author} {\bibfnamefont {J.}~\bibnamefont
  {Zhou}},\ }\bibfield  {title} {\bibinfo {title} {{Transverse single-spin
  asymmetry in direct photon production in polarized pA collisions}},\ }\href
  {https://doi.org/10.1103/PhysRevD.90.034016} {\bibfield  {journal} {\bibinfo
  {journal} {Phys. Rev. D}\ }\textbf {\bibinfo {volume} {90}},\ \bibinfo
  {pages} {034016} (\bibinfo {year} {2014})}\BibitemShut {NoStop}%
\bibitem [{\citenamefont {Zhou}(2015)}]{Zhou:2015ima}%
  \BibitemOpen
  \bibfield  {author} {\bibinfo {author} {\bibfnamefont {J.}~\bibnamefont
  {Zhou}},\ }\bibfield  {title} {\bibinfo {title} {{Transverse single-spin
  asymmetry in Drell-Yan production in polarized pA collisions}},\ }\href
  {https://doi.org/10.1103/PhysRevD.92.014034} {\bibfield  {journal} {\bibinfo
  {journal} {Phys. Rev. D}\ }\textbf {\bibinfo {volume} {92}},\ \bibinfo
  {pages} {014034} (\bibinfo {year} {2015})}\BibitemShut {NoStop}%
\bibitem [{\citenamefont {Hatta}\ \emph {et~al.}(2016)\citenamefont {Hatta},
  \citenamefont {Xiao}, \citenamefont {Yoshida},\ and\ \citenamefont
  {Yuan}}]{Hatta:2016wjz}%
  \BibitemOpen
  \bibfield  {author} {\bibinfo {author} {\bibfnamefont {Y.}~\bibnamefont
  {Hatta}}, \bibinfo {author} {\bibfnamefont {B.~W.}\ \bibnamefont {Xiao}},
  \bibinfo {author} {\bibfnamefont {S.}~\bibnamefont {Yoshida}},\ and\ \bibinfo
  {author} {\bibfnamefont {F.}~\bibnamefont {Yuan}},\ }\bibfield  {title}
  {\bibinfo {title} {{Single-Spin Asymmetry in Forward $pA$ Collisions}},\
  }\href {https://doi.org/10.1103/PhysRevD.94.054013} {\bibfield  {journal}
  {\bibinfo  {journal} {Phys. Rev. D}\ }\textbf {\bibinfo {volume} {94}},\
  \bibinfo {pages} {054013} (\bibinfo {year} {2016})}\BibitemShut {NoStop}%
\bibitem [{\citenamefont {Hatta}\ \emph {et~al.}(2017)\citenamefont {Hatta},
  \citenamefont {Xiao}, \citenamefont {Yoshida},\ and\ \citenamefont
  {Yuan}}]{Hatta:2016khv}%
  \BibitemOpen
  \bibfield  {author} {\bibinfo {author} {\bibfnamefont {Y.}~\bibnamefont
  {Hatta}}, \bibinfo {author} {\bibfnamefont {B.~W.}\ \bibnamefont {Xiao}},
  \bibinfo {author} {\bibfnamefont {S.}~\bibnamefont {Yoshida}},\ and\ \bibinfo
  {author} {\bibfnamefont {F.}~\bibnamefont {Yuan}},\ }\bibfield  {title}
  {\bibinfo {title} {{Single-spin asymmetry in forward $pA$ collisions II:
  Fragmentation contribution}},\ }\href
  {https://doi.org/10.1103/PhysRevD.95.014008} {\bibfield  {journal} {\bibinfo
  {journal} {Phys. Rev. D}\ }\textbf {\bibinfo {volume} {95}},\ \bibinfo
  {pages} {014008} (\bibinfo {year} {2017})}\BibitemShut {NoStop}%
\bibitem [{\citenamefont {Aidala}\ \emph {et~al.}(2019)\citenamefont {Aidala}
  \emph {et~al.}}]{PHENIX:2019ouo}%
  \BibitemOpen
  \bibfield  {author} {\bibinfo {author} {\bibfnamefont {C.}~\bibnamefont
  {Aidala}} \emph {et~al.} (\bibinfo {collaboration} {PHENIX Collaboration}),\
  }\bibfield  {title} {\bibinfo {title} {{Nuclear Dependence of the Transverse
  Single-Spin Asymmetry in the Production of Charged Hadrons at Forward
  Rapidity in Polarized $p+p$, $p+$Al, and $p+$Au Collisions at
  $\sqrt{s_{_{NN}}}=200$ GeV}},\ }\href
  {https://doi.org/10.1103/PhysRevLett.123.122001} {\bibfield  {journal}
  {\bibinfo  {journal} {Phys. Rev. Lett.}\ }\textbf {\bibinfo {volume} {123}},\
  \bibinfo {pages} {122001} (\bibinfo {year} {2019})}\BibitemShut {NoStop}%
\bibitem [{\citenamefont {Adam}\ \emph
  {et~al.}(2021{\natexlab{b}})\citenamefont {Adam} \emph
  {et~al.}}]{STAR:2020grs}%
  \BibitemOpen
  \bibfield  {author} {\bibinfo {author} {\bibfnamefont {J.}~\bibnamefont
  {Adam}} \emph {et~al.} (\bibinfo {collaboration} {STAR Collaboration}),\
  }\bibfield  {title} {\bibinfo {title} {{Comparison of transverse single-spin
  asymmetries for forward $\pi^{0}$ production in polarized $pp$, $p\rm{Al}$
  and $p\rm{Au}$ collisions at nucleon pair c.m. energy
  $\sqrt{s_{\mathrm{NN}}}= 200$ GeV}},\ }\href
  {https://doi.org/10.1103/PhysRevD.103.072005} {\bibfield  {journal} {\bibinfo
   {journal} {Phys. Rev. D}\ }\textbf {\bibinfo {volume} {103}},\ \bibinfo
  {pages} {072005} (\bibinfo {year} {2021}{\natexlab{b}})}\BibitemShut
  {NoStop}%
\bibitem [{\citenamefont {Adcox}\ \emph {et~al.}(2003)\citenamefont {Adcox}
  \emph {et~al.}}]{PHENIX:2003nhg}%
  \BibitemOpen
  \bibfield  {author} {\bibinfo {author} {\bibfnamefont {K.}~\bibnamefont
  {Adcox}} \emph {et~al.} (\bibinfo {collaboration} {PHENIX Collaboration}),\
  }\bibfield  {title} {\bibinfo {title} {{PHENIX detector overview}},\ }\href
  {https://doi.org/10.1016/S0168-9002(02)01950-2} {\bibfield  {journal}
  {\bibinfo  {journal} {Nucl. Instrum. Methods Phys. Res., Sec. A}\ }\textbf
  {\bibinfo {volume} {499}},\ \bibinfo {pages} {469} (\bibinfo {year}
  {2003})}\BibitemShut {NoStop}%
\bibitem [{\citenamefont {Akikawa}\ \emph {et~al.}(2003)\citenamefont {Akikawa}
  \emph {et~al.}}]{PHENIX:2003yhi}%
  \BibitemOpen
  \bibfield  {author} {\bibinfo {author} {\bibfnamefont {H.}~\bibnamefont
  {Akikawa}} \emph {et~al.} (\bibinfo {collaboration} {PHENIX Collaboration}),\
  }\bibfield  {title} {\bibinfo {title} {{PHENIX muon arms}},\ }\href
  {https://doi.org/10.1016/S0168-9002(02)01955-1} {\bibfield  {journal}
  {\bibinfo  {journal} {Nucl. Instrum. Methods Phys. Res., Sec. A}\ }\textbf
  {\bibinfo {volume} {499}},\ \bibinfo {pages} {537} (\bibinfo {year}
  {2003})}\BibitemShut {NoStop}%
\bibitem [{\citenamefont {Adare}\ \emph {et~al.}(2012)\citenamefont {Adare}
  \emph {et~al.}}]{PHENIX:2012itj}%
  \BibitemOpen
  \bibfield  {author} {\bibinfo {author} {\bibfnamefont {A.}~\bibnamefont
  {Adare}} \emph {et~al.} (\bibinfo {collaboration} {PHENIX Collaboration}),\
  }\bibfield  {title} {\bibinfo {title} {{Nuclear-Modification Factor for
  Open-Heavy-Flavor Production at Forward Rapidity in Cu+Cu Collisions at
  $\sqrt{s_{NN}}=200$ GeV}},\ }\href
  {https://doi.org/10.1103/PhysRevC.86.024909} {\bibfield  {journal} {\bibinfo
  {journal} {Phys. Rev. C}\ }\textbf {\bibinfo {volume} {86}},\ \bibinfo
  {pages} {024909} (\bibinfo {year} {2012})}\BibitemShut {NoStop}%
\bibitem [{\citenamefont {Allen}\ \emph {et~al.}(2003)\citenamefont {Allen}
  \emph {et~al.}}]{PHENIX:2003tlh}%
  \BibitemOpen
  \bibfield  {author} {\bibinfo {author} {\bibfnamefont {M.}~\bibnamefont
  {Allen}} \emph {et~al.} (\bibinfo {collaboration} {PHENIX Collaboration}),\
  }\bibfield  {title} {\bibinfo {title} {{PHENIX inner detectors}},\ }\href
  {https://doi.org/10.1016/S0168-9002(02)01956-3} {\bibfield  {journal}
  {\bibinfo  {journal} {Nucl. Instrum. Methods Phys. Res., Sec. A}\ }\textbf
  {\bibinfo {volume} {499}},\ \bibinfo {pages} {549} (\bibinfo {year}
  {2003})}\BibitemShut {NoStop}%
\bibitem [{\citenamefont {Adachi}\ \emph {et~al.}(2013)\citenamefont {Adachi}
  \emph {et~al.}}]{Adachi:2013qha}%
  \BibitemOpen
  \bibfield  {author} {\bibinfo {author} {\bibfnamefont {S.}~\bibnamefont
  {Adachi}} \emph {et~al.},\ }\bibfield  {title} {\bibinfo {title} {{Trigger
  electronics upgrade of PHENIX muon tracker}},\ }\href
  {https://doi.org/10.1016/j.nima.2012.11.088} {\bibfield  {journal} {\bibinfo
  {journal} {Nucl. Instrum. Methods Phys. Res., Sec. A}\ }\textbf {\bibinfo
  {volume} {703}},\ \bibinfo {pages} {114} (\bibinfo {year}
  {2013})}\BibitemShut {NoStop}%
\bibitem [{\citenamefont {{RHIC Polarimetry Group}}()}]{polarimetry}%
  \BibitemOpen
  \bibfield  {author} {\bibinfo {author} {\bibnamefont {{RHIC Polarimetry
  Group}}},\ }\bibfield  {title} {\bibinfo {title} {{RHIC polarization for Runs
  9--12}},\ }\bibinfo {note} {{RHIC/CAD Accelerator Physics Note 490
  (2018)}}\BibitemShut {NoStop}%
\bibitem [{\citenamefont {Aidala}\ \emph {et~al.}(2020)\citenamefont {Aidala}
  \emph {et~al.}}]{PHENIX:2019gix}%
  \BibitemOpen
  \bibfield  {author} {\bibinfo {author} {\bibfnamefont {C.}~\bibnamefont
  {Aidala}} \emph {et~al.} (\bibinfo {collaboration} {PHENIX Collaboration}),\
  }\bibfield  {title} {\bibinfo {title} {{Nuclear-modification factor of
  charged hadrons at forward and backward rapidity in $p+$Al and $p+$Au
  collisions at $\sqrt{s_{_{NN}}}=200$ GeV}},\ }\href
  {https://doi.org/10.1103/PhysRevC.101.034910} {\bibfield  {journal} {\bibinfo
   {journal} {Phys. Rev. C}\ }\textbf {\bibinfo {volume} {101}},\ \bibinfo
  {pages} {034910} (\bibinfo {year} {2020})}\BibitemShut {NoStop}%
\bibitem [{\citenamefont {Adare}\ \emph
  {et~al.}(2014{\natexlab{c}})\citenamefont {Adare} \emph
  {et~al.}}]{PHENIX:2013txu}%
  \BibitemOpen
  \bibfield  {author} {\bibinfo {author} {\bibfnamefont {A.}~\bibnamefont
  {Adare}} \emph {et~al.} (\bibinfo {collaboration} {PHENIX Collaboration}),\
  }\bibfield  {title} {\bibinfo {title} {{Cold-Nuclear-Matter Effects on
  Heavy-Quark Production at Forward and Backward Rapidity in d+Au Collisions at
  $\sqrt{s_{NN}}=200$ GeV}},\ }\href
  {https://doi.org/10.1103/PhysRevLett.112.252301} {\bibfield  {journal}
  {\bibinfo  {journal} {Phys. Rev. Lett.}\ }\textbf {\bibinfo {volume} {112}},\
  \bibinfo {pages} {252301} (\bibinfo {year} {2014}{\natexlab{c}})}\BibitemShut
  {NoStop}%
\bibitem [{\citenamefont {Sjostrand}\ \emph {et~al.}()\citenamefont
  {Sjostrand}, \citenamefont {Mrenna},\ and\ \citenamefont
  {Skands}}]{Sjostrand:2006za}%
  \BibitemOpen
  \bibfield  {author} {\bibinfo {author} {\bibfnamefont {T.}~\bibnamefont
  {Sjostrand}}, \bibinfo {author} {\bibfnamefont {S.}~\bibnamefont {Mrenna}},\
  and\ \bibinfo {author} {\bibfnamefont {P.~Z.}\ \bibnamefont {Skands}},\
  }\href {https://doi.org/10.1088/1126-6708/2006/05/026} {\bibinfo {title}
  {{PYTHIA 6.4 Physics and Manual}}},\ \bibinfo {note} {{J. High Energy Phys.
  {\bf 05 (2006)} 026}}\BibitemShut {NoStop}%
\bibitem [{\citenamefont {Gyulassy}\ and\ \citenamefont
  {Wang}(1994)}]{Gyulassy:1994ew}%
  \BibitemOpen
  \bibfield  {author} {\bibinfo {author} {\bibfnamefont {M.}~\bibnamefont
  {Gyulassy}}\ and\ \bibinfo {author} {\bibfnamefont {X.}~\bibnamefont
  {Wang}},\ }\bibfield  {title} {\bibinfo {title} {{HIJING 1.0: A Monte Carlo
  program for parton and particle production in high-energy hadronic and
  nuclear collisions}},\ }\href {https://doi.org/10.1016/0010-4655(94)90057-4}
  {\bibfield  {journal} {\bibinfo  {journal} {Comput. Phys. Commun.}\ }\textbf
  {\bibinfo {volume} {83}},\ \bibinfo {pages} {307} (\bibinfo {year}
  {1994})}\BibitemShut {NoStop}%
\bibitem [{\citenamefont {Adare}\ \emph {et~al.}(2011)\citenamefont {Adare}
  \emph {et~al.}}]{PHENIX:2011rvu}%
  \BibitemOpen
  \bibfield  {author} {\bibinfo {author} {\bibfnamefont {A.}~\bibnamefont
  {Adare}} \emph {et~al.} (\bibinfo {collaboration} {PHENIX Collaboration}),\
  }\bibfield  {title} {\bibinfo {title} {{Identified charged hadron production
  in $p+p$ collisions at $\sqrt{s}=200$ and 62.4 GeV}},\ }\href
  {https://doi.org/10.1103/PhysRevC.83.064903} {\bibfield  {journal} {\bibinfo
  {journal} {Phys. Rev. C}\ }\textbf {\bibinfo {volume} {83}},\ \bibinfo
  {pages} {064903} (\bibinfo {year} {2011})}\BibitemShut {NoStop}%
\bibitem [{\citenamefont {Agakishiev}\ \emph {et~al.}(2012)\citenamefont
  {Agakishiev} \emph {et~al.}}]{STAR:2011iap}%
  \BibitemOpen
  \bibfield  {author} {\bibinfo {author} {\bibfnamefont {G.}~\bibnamefont
  {Agakishiev}} \emph {et~al.} (\bibinfo {collaboration} {STAR
  Collaboration}),\ }\bibfield  {title} {\bibinfo {title} {{Identified hadron
  compositions in p+p and Au+Au collisions at high transverse momenta at
  $\sqrt{s_{_{NN}}} = 200$ GeV}},\ }\href
  {https://doi.org/10.1103/PhysRevLett.108.072302} {\bibfield  {journal}
  {\bibinfo  {journal} {Phys. Rev. Lett.}\ }\textbf {\bibinfo {volume} {108}},\
  \bibinfo {pages} {072302} (\bibinfo {year} {2012})}\BibitemShut {NoStop}%
\bibitem [{\citenamefont {Adare}\ \emph {et~al.}(2013)\citenamefont {Adare}
  \emph {et~al.}}]{PHENIX:2013kod}%
  \BibitemOpen
  \bibfield  {author} {\bibinfo {author} {\bibfnamefont {A.}~\bibnamefont
  {Adare}} \emph {et~al.} (\bibinfo {collaboration} {PHENIX Collaboration}),\
  }\bibfield  {title} {\bibinfo {title} {{Spectra and ratios of identified
  particles in Au+Au and $d$+Au collisions at $\sqrt{s_{NN}}=200$ GeV}},\
  }\href {https://doi.org/10.1103/PhysRevC.88.024906} {\bibfield  {journal}
  {\bibinfo  {journal} {Phys. Rev. C}\ }\textbf {\bibinfo {volume} {88}},\
  \bibinfo {pages} {024906} (\bibinfo {year} {2013})}\BibitemShut {NoStop}%
\bibitem [{\citenamefont {Agostinelli}\ \emph {et~al.}(2003)\citenamefont
  {Agostinelli} \emph {et~al.}}]{GEANT4:2002zbu}%
  \BibitemOpen
  \bibfield  {author} {\bibinfo {author} {\bibfnamefont {S.}~\bibnamefont
  {Agostinelli}} \emph {et~al.} (\bibinfo {collaboration} {GEANT4
  Collaboration}),\ }\bibfield  {title} {\bibinfo {title} {{GEANT4--a
  simulation toolkit}},\ }\href {https://doi.org/10.1016/S0168-9002(03)01368-8}
  {\bibfield  {journal} {\bibinfo  {journal} {Nucl. Instrum. Methods Phys.
  Res., Sec. A}\ }\textbf {\bibinfo {volume} {506}},\ \bibinfo {pages} {250}
  (\bibinfo {year} {2003})}\BibitemShut {NoStop}%
\bibitem [{\citenamefont {Workman}\ \emph {et~al.}(2022)\citenamefont {Workman}
  \emph {et~al.}}]{Workman:2022ynf}%
  \BibitemOpen
  \bibfield  {author} {\bibinfo {author} {\bibfnamefont {R.~L.}\ \bibnamefont
  {Workman}} \emph {et~al.} (\bibinfo {collaboration} {Particle Data Group}),\
  }\bibfield  {title} {\bibinfo {title} {{Review of Particle Physics}},\ }\href
  {https://doi.org/10.1093/ptep/ptac097} {\bibfield  {journal} {\bibinfo
  {journal} {Prog. Theor. Exp. Phys.}\ }\textbf {\bibinfo {volume} {2022}},\
  \bibinfo {pages} {083C01} (\bibinfo {year} {2022})}\BibitemShut {NoStop}%
\bibitem [{\citenamefont {Allison}\ \emph {et~al.}(2016)\citenamefont {Allison}
  \emph {et~al.}}]{Allison:2016lfl}%
  \BibitemOpen
  \bibfield  {author} {\bibinfo {author} {\bibfnamefont {J.}~\bibnamefont
  {Allison}} \emph {et~al.},\ }\bibfield  {title} {\bibinfo {title} {{Recent
  developments in Geant4}},\ }\href
  {https://doi.org/10.1016/j.nima.2016.06.125} {\bibfield  {journal} {\bibinfo
  {journal} {Nucl. Instrum. Methods Phys. Res., Sec. A}\ }\textbf {\bibinfo
  {volume} {835}},\ \bibinfo {pages} {186} (\bibinfo {year}
  {2016})}\BibitemShut {NoStop}%
\bibitem [{\citenamefont {Aidala}\ \emph {et~al.}(2017)\citenamefont {Aidala}
  \emph {et~al.}}]{PHENIX:2017wbv}%
  \BibitemOpen
  \bibfield  {author} {\bibinfo {author} {\bibfnamefont {C.}~\bibnamefont
  {Aidala}} \emph {et~al.} (\bibinfo {collaboration} {PHENIX Collaboration}),\
  }\bibfield  {title} {\bibinfo {title} {{Cross section and transverse
  single-spin asymmetry of muons from open heavy-flavor decays in polarized
  $p$+$p$ collisions at $\sqrt{s}=200$ GeV}},\ }\href
  {https://doi.org/10.1103/PhysRevD.95.112001} {\bibfield  {journal} {\bibinfo
  {journal} {Phys. Rev. D}\ }\textbf {\bibinfo {volume} {95}},\ \bibinfo
  {pages} {112001} (\bibinfo {year} {2017})}\BibitemShut {NoStop}%
\bibitem [{\citenamefont {Aidala}\ \emph {et~al.}(2018)\citenamefont {Aidala}
  \emph {et~al.}}]{PHENIX:2018qvl}%
  \BibitemOpen
  \bibfield  {author} {\bibinfo {author} {\bibfnamefont {C.}~\bibnamefont
  {Aidala}} \emph {et~al.} (\bibinfo {collaboration} {PHENIX Collaboration}),\
  }\bibfield  {title} {\bibinfo {title} {{Single-spin asymmetry of $J/\psi$
  production in $p+p$, $p+$Al, and $p+$Au collisions with transversely
  polarized proton beams at $\sqrt{s_{_{NN}}}=200$ GeV}},\ }\href
  {https://doi.org/10.1103/PhysRevD.98.012006} {\bibfield  {journal} {\bibinfo
  {journal} {Phys. Rev. D}\ }\textbf {\bibinfo {volume} {98}},\ \bibinfo
  {pages} {012006} (\bibinfo {year} {2018})}\BibitemShut {NoStop}%
\bibitem [{\citenamefont {Ohlsen}\ and\ \citenamefont
  {Keaton}(1973)}]{Ohlsen:1973wf}%
  \BibitemOpen
  \bibfield  {author} {\bibinfo {author} {\bibfnamefont {G.~G.}\ \bibnamefont
  {Ohlsen}}\ and\ \bibinfo {author} {\bibfnamefont {P.~W.}\ \bibnamefont
  {Keaton}},\ }\bibfield  {title} {\bibinfo {title} {{Techniques for
  measurement of spin-1/2 and spin-1 polarization analyzing tensors}},\ }\href
  {https://doi.org/10.1016/0029-554X(73)90450-3} {\bibfield  {journal}
  {\bibinfo  {journal} {Nucl. Instrum. Methods Phys. Res., Sec. A}\ }\textbf
  {\bibinfo {volume} {109}},\ \bibinfo {pages} {41} (\bibinfo {year}
  {1973})}\BibitemShut {NoStop}%
\bibitem [{\citenamefont {Zhou}(2017)}]{Zhou:2017sdx}%
  \BibitemOpen
  \bibfield  {author} {\bibinfo {author} {\bibfnamefont {J.}~\bibnamefont
  {Zhou}},\ }\bibfield  {title} {\bibinfo {title} {{Single-spin asymmetries in
  forward p-p/A collisions revisited: the role of color entanglement}},\ }\href
  {https://doi.org/10.1103/PhysRevD.96.034027} {\bibfield  {journal} {\bibinfo
  {journal} {Phys. Rev. D}\ }\textbf {\bibinfo {volume} {96}},\ \bibinfo
  {pages} {034027} (\bibinfo {year} {2017})}\BibitemShut {NoStop}%
\bibitem [{\citenamefont {Beni\'c}\ and\ \citenamefont
  {Hatta}(2019)}]{Benic:2018amn}%
  \BibitemOpen
  \bibfield  {author} {\bibinfo {author} {\bibfnamefont {S.}~\bibnamefont
  {Beni\'c}}\ and\ \bibinfo {author} {\bibfnamefont {Y.}~\bibnamefont
  {Hatta}},\ }\bibfield  {title} {\bibinfo {title} {{Single-spin asymmetry in
  forward $pA$ collisions: Phenomenology at RHIC}},\ }\href
  {https://doi.org/10.1103/PhysRevD.99.094012} {\bibfield  {journal} {\bibinfo
  {journal} {Phys. Rev. D}\ }\textbf {\bibinfo {volume} {99}},\ \bibinfo
  {pages} {094012} (\bibinfo {year} {2019})}\BibitemShut {NoStop}%
\bibitem [{\citenamefont {Kovchegov}\ and\ \citenamefont
  {Santiago}(2020)}]{Kovchegov:2020kxg}%
  \BibitemOpen
  \bibfield  {author} {\bibinfo {author} {\bibfnamefont {Y.~V.}\ \bibnamefont
  {Kovchegov}}\ and\ \bibinfo {author} {\bibfnamefont {M.~G.}\ \bibnamefont
  {Santiago}},\ }\bibfield  {title} {\bibinfo {title} {{Lensing mechanism meets
  small- $x$ physics: Single transverse spin asymmetry in $p^{\uparrow}+p$ and
  $p^{\uparrow}+A$ collisions}},\ }\href
  {https://doi.org/10.1103/PhysRevD.102.014022} {\bibfield  {journal} {\bibinfo
   {journal} {Phys. Rev. D}\ }\textbf {\bibinfo {volume} {102}},\ \bibinfo
  {pages} {014022} (\bibinfo {year} {2020})}\BibitemShut {NoStop}%
\end{thebibliography}

%
 
\end{document}